**Highlights**

- Developed a dedicated software tool for generation and prediction ("hacking") of LFSR and Mersenne Twister pseudo-random sequences.
- Two operational modes: random sequence generation and estimation/hacking of arbitrary input streams.
- Implemented statistical and algebraic estimation techniques to reconstruct LFSR polynomials and MT internal states with high accuracy.
- Demonstrated the use of deep learning to predict MT sequences, achieving near-perfect accuracy by exploiting deterministic correlations.
- Hybrid LFSR–MT pseudo-random generators: showing increased complexity but persistent vulnerability to prediction.
- Verified the resilience of quantum random number sequences, which remained unbreakable under all applied prediction methods.
- Confirms the superiority of quantum random sequences, offering true non-deterministic security compared to classical PRNGs.
- Propose robust quantum jamming based on using quantum random numbers as a quantum secure alternative to classical frequency-hopping countermeasures.

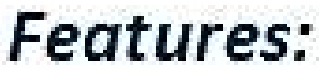

*Features:*

- *Prediction LFSR and Mersenne Twister (MT) without Unambiguous*
- *Hybrid LFSR-MT sequences Prediction*
- *Analysis Quantum random number sequences*
- *Deep learning based Estimation MT*
- *Generation and Analysis Random sequences*

Classical PRNGs

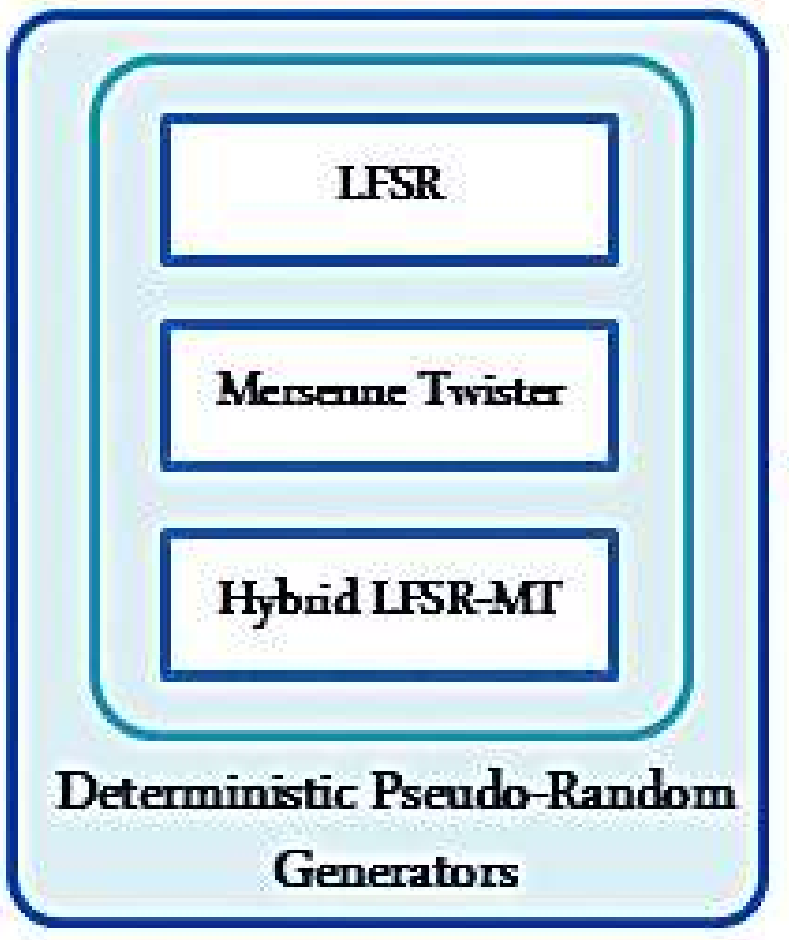

Quantum Random Sequences

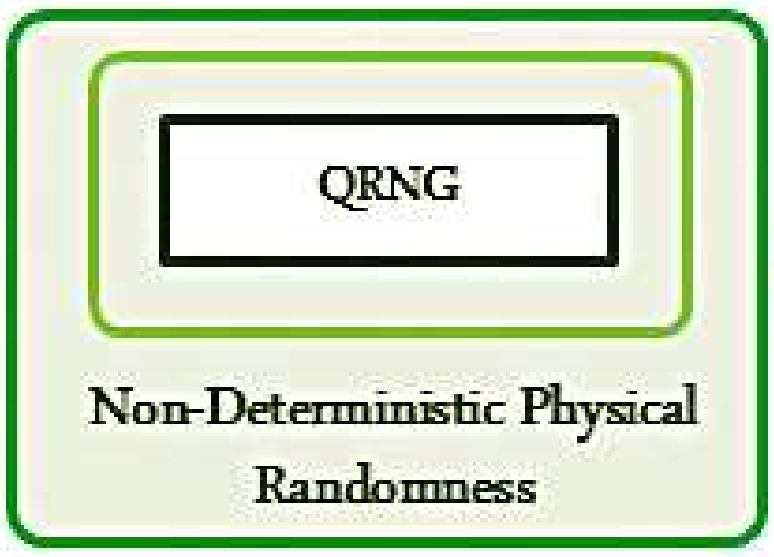

Software Platform & GUI

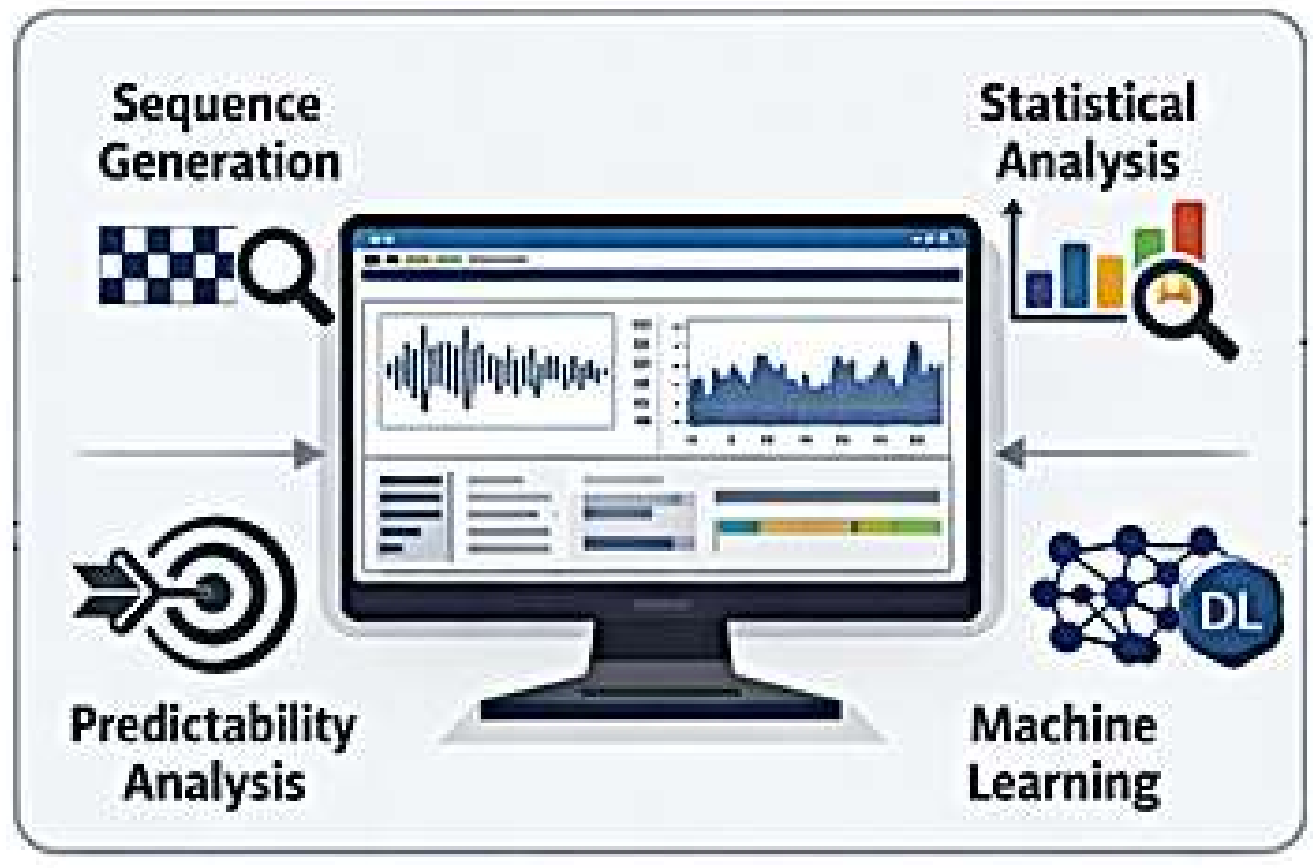

*Article*

# Software Platform for Hybrid Pseudo-Random Sequence Generation and Predictability Analysis Based on LFSR and Mersenne Twister

**Ali Abdolrahimi Zarnagh [1, 2] and Ali Motazedifard [3, 1, *]**

[1] *Quantum Remote Sensing Lab, Quantum Metrology Group, Iranian Center for Quantum Technologies (ICQT), Tehran, Tehran 15998-14713, Iran*
[2] *School of Electrical and Computer Engineering, University of Tehran, North Kargar Ave., Tehran 14395-515, Iran*
[3] *Department of Physics, University of Tehran, Kargar Shomali Ave, Tehran 14399-55961, Iran*

[*] Correspondence: alimotazedifard@ut.ac.ir & motazedifard.ali@gmail.com

**Abstract**

Generating high-quality random and pseudo-random sequences is essential for a wide range of electronic and signal-processing applications, including secure communications, radar systems, spread-spectrum techniques, and autonomous platforms. While true and quantum random number generators (QRNGs) offer superior unpredictability, classical pseudo-random number generators (PRNGs) such as Linear Feedback Shift Registers (LFSRs) and the Mersenne Twister (MT) remain widely used due to their computational efficiency and ease of implementation. In this work, we present a user-friendly software platform designed for the generation, analysis, and predictability assessment of pseudo-random bit sequences. The developed software operates in two main modes: (i) generation of arbitrary pseudo-random sequences based on classical PRNGs and their hybrid combinations, and (ii) estimation and predictability analysis of input pseudo-random sequences using statistical metrics and data-driven approaches. Hybrid configurations combining LFSR and MT are investigated to evaluate their impact on sequence complexity and resistance to prediction. Furthermore, the platform integrates machine-learning-based analysis tools, including a deep learning framework, to study the conditions under which deterministic PRNGs may remain partially predictable despite increased structural complexity. The obtained results highlight inherent limitations associated with algorithmic random sequence generators and motivate the use of quantum random sequences in security-critical scenarios. Using the proposed software, a comparative evaluation between classical LFSR–MT sequences and quantum random sequences is performed, demonstrating the enhanced unpredictability of quantum randomness arising from its non-deterministic physical origin. As an application outlook, the use of quantum random sequences in a jamming framework and its improved robustness against predictability-based attacks are discussed. Overall, the proposed software serves as a practical analysis and benchmarking tool for the design and evaluation of random sequence generators in emerging electronic, sensing, and quantum-enabled communication systems.

## 1. Introduction

Random sequences (RS) have been applied in various domains, including cryptography, secure radars with phase- or amplitude-coded signals, spread spectrum communications, and autonomous vehicle systems [1-4]. The most important applications of RSs in cryptography include key generation, initialization vectors (IVs) production for block ciphers, generating pseudo-random sequences with plain text in stream ciphers, generating random challenges in zero-knowledge proofs, and more [5-7].

Additionally, random numbers (RNs) in radar systems are used for pulse compression to increase resolution and target detection, further signal randomization to counter jamming [8], clutter rejection techniques to improve detection in noisy environments, generating noise signals in radar for better detection and lower tracking probability [9], and other purposes.

In spread spectrum communications, the most important applications of RNs include generating hopping patterns in frequency hopping spread spectrum (FHSS), creating random sequences for spectral spreading in direct-sequence spread spectrum (DSSS), assigning unique codes to each user for bandwidth usage in code-division multiple access (CDMA) methods, and other techniques [10-11]. Furthermore, more recently random numbers are used in autonomous vehicles for modeling noise received by lidars, radars, and vehicle cameras, modeling and simulating noisy environments, urban traffic modeling, and simulating movements of other environmental factors like pedestrians [12-15].

However, generating truly random numbers in the real world is challenging because accessing and controlling random phenomena is expensive and often not cost-effective for many applications [16]. Therefore, as an alternative solution, pseudo-random generators are used [17]. Pseudo-random generators do not inherently produce random numbers but use a deterministic and precise algorithm to generate pseudo-random numbers (PRNs) [18]. These pseudo-random algorithms are used as alternatives to true random number generators (TRNGs) due to their speed and ease of producing numbers with suitable random quality, while they are not inherently random [19].

Pseudo-random number generators (PRNG) use certain and definable structures, making them ultimately predictable [20]. However, the computational complexity of breaking their structure is difficult, such that it is not feasible through classical computational systems within a reasonable timeframe [21].

PRNG algorithms are specific and can be expressed through algebraic relationships, making them predictable [22]. If their output streams are examined over long periods, one can conclude their repeatable characteristics [23], and therefore, predicting patterns is a definitive feature of them. However, examining long periods for classical systems is challenging [18].

Recent advancements in quantum computing technology have led to concerns that quantum computers could be used to hack and predict classical random numbers [24]. On the other hands, QRNGs have inherent randomness which do not leave traces even over long periods because they are not the result of a repeatable algorithm [16,25,26]. This feature is originated to the quantum vacuum fluctuation.

One of the most important structures used for generating pseudo-random bit streams is the Linear Feedback Shifted Registers (LFSR) structure [27]. LFSRs are used in many industrial, laboratory, and commercial applications. Their appropriate speed and low memory requirements contribute to their popularity [28]. The LFSR structure is applied in various systems like stream cipher cryptography, spread spectrum methods, and secure radar systems [29,30]. Another popular pseudo-random number generator is the Mersenne-Twister (MT) algorithm, which has been used in many software applications like MATLAB. The MT structure also employs a twisted LFSR structure. On the output of the twisted LFSR, which represents the state vector, algebraic functions are used to bring the final output closer to a uniform distribution [31]. The structures of the introduced pseudo-random algorithms are breakable due to pattern generation in specific periods.

To predict and estimate the continuation of sequences produced by pseudo-random generators, their algorithmic structure must be broken [32]. For LFSR, this is done by estimating the generator polynomial and initial state [29]. The introduced method (based on the premise that it has a roughly uniform distribution of zero and one bits) uses a statistical test to estimate the generator polynomial and initial state. Therefore, in this case, the estimation method is statistical and not deterministic. For the MT algorithm, structure estimation is also performed by discovering the state vector. In the MT method, with access to the state vector, the output of the MT algorithm can be reproduced. Consequently, in this method, we use a completely deterministic structure for estimation. Since the MT output number is the result of applying a series of deterministic algebraic operations on the state vector, an artificial intelligence (AI) model can be used to learn this function. In this case, breaking the MT structure using an AI method is also possible. In our developed software, a deep structure for breaking the MT algorithm has been trained that has a 100% accuracy in determining subsequent states in the algorithm [33].

Pseudo-random algorithms leave a deterministic effect on specific periods in the bit streams produced by pseudo-random generators [23]. By examining these bit streams in periods with a high number of iterations, these effects can be detected, but as mentioned, this is computationally impossible within a limited timeframe through classical computers. While in the future, a quantum computer will perform this task in a practical timeframe, as it is far faster than a classical computer. Therefore, generally, the periodicity characteristic in pseudo-random generators is not the result of true randomness, and this randomness is applied to the system through complexity [17].

To demonstrate that the presence of a pattern causes the breakdown of the introduced pseudo-random structures, the observation is used that such patterns never exist in truly quantum random data. As a result, quantum random numbers are not predictable because they are inherently random with no behind algorithm. In contrast, pseudo-random numbers are hackable due to the predictable patterns in their generated bit streams.

Motivated by the above investigations, we developed a software to hack and estimate pseudo-random streams from LFSR and MT generators. This is achieved by using deterministic, classical, and AI-based statistical methods that exploit the algorithmic patterns in the generated bit streams. The software features two modes: one for generating a bit stream of an arbitrary length and method, and a second for estimating or hacking an arbitrary input stream. It enables to break the polynomial and initial state of the LFSR from an input bit stream with a uniform distribution of 0 and 1. Additionally, the MT algorithm generator (which has a structure based on LFSR) can be examined in the software, and by analyzing the state vector of the MT algorithm, its structure is broken. Also, a trained neural network is used in our GUI/software to determine the state vector of the MT algorithm [34].

Finally, users can get a report that includes the processing time and match points for different input data with varying bit lengths. The software also provides the ability to test quantum random number streams to ensure they cannot be hacked or estimated. This implies that, unlike PRNs, there is no underlying algorithm behind QRNs. Additionally, we briefly introduce an improved type of quantum jamming that exploits QRNs to be robust against attacks, unlike classical jamming, which can be identified and estimated. Lastly, it should be pointed out that the developed software can be used as a supportive tool in many quantum sensing and communication devices.

The structural organization of the paper is as follows. The **Section. 2** presents fundamental discussions concerning randomness theory and its conceptual foundations. Subsequently, the **Section**. **3** elucidates the scientific methodology, generator architectures, and analytical estimations implemented in the developed software. Then, they are applied to hack the LFSR and MT streams. The developed GUI software is introduced and tested in **Section. 4**. Some results are presented in this section. Finally, the conclusion remarks, discussion, new perspectives, and outlooks are presented in **Section. 5**.

## 2. Background and Concepts

In this section, the theory and concepts are briefly presented.

### *2.1 Degree of Randomness*

#### 2.1.1 Concept of Randomness

A random bit stream consists of a sequence of numbers generated through independent, unbiased processes, such as idealized fair coin tosses where each outcome (0 or 1) occurs with equal probability (0.5). However, it is obvious that this is a statistical randomness. It originates from our inability to exactly solve the mechanical equations of motion in real-world conditions, not from an inherent or fundamental random process like quantum mechanical randomness. In such a stream, the results of consecutive tosses are entirely independent, ensuring no member of the sequence influences subsequent values, thereby rendering future bits consequently are unpredictable. This conceptual framework positions the fair coin toss as a theoretical generator of perfectly classical random bit streams, characterized by uniform distribution and statistical independence [35].

However, while coin tossing serves as a metaphor for randomness, it is not inherently random in practice. The dynamics of a physical coin toss involve deterministic forces, but its complexity defies precise modeling, leading engineers to treat it as a random process for practical purposes. For example, in practical problems of telecommunications, phase calculations in non-coherent channels are often approximated as random variables due to measurement limitations, sacrificing precision to manage computational complexity [1]. Such compromises underscore the challenge of accessing true randomness in real-world systems, prompting reliance on pseudo-random numbers—deterministic sequences that mimic randomness through algorithmic complexity.

In contrast, quantum processes generate inherently random sequences rooted in the fundamental principles of quantum mechanics. Heisenberg's uncertainty principle dictates that precise simultaneous measurement of complementary parameters (e.g., momentum- and position-quadrature) is impossible; increasing precision in one increases the uncertainty in the other [36]. This intrinsic ambiguity, generalizable to all quantum phenomena, equates to true randomness, independent of measurement limitations or computational constraints [37]. The rejection of hidden variable theories, validated by Bell's theorems [38], further confirms that quantum randomness arises from the non-deterministic nature of quantum systems, not observational shortcomings. This can be attributed, for example, to quantum vacuum fluctuations or the quantum-state superposition principle, both of which stem from the second quantization in quantum theory.

Though idealized coin tosses remain impractical for actual random bit generation, they establish a theoretical benchmark for evaluating both true random number generators (e.g., quantum-based systems) and pseudo-random algorithms [35]. This duality highlights the gap between conceptual models of randomness and practical implementations, emphasizing the need for robust frameworks to assess and validate randomness in cryptographic, engineering, and scientific applications.

#### 2.1.2 Unpredictability

Random and Pseudo-random numbers produced must be unpredictable [7]. For pseudo-random number generators (PRNG), if the Seed is not accessible to any user, despite having previously generated random numbers in the stream, the next output in the stream should be unpredictable [21]. Generally, no connection between a Seed and its generated value should be extractable. Each stream member should be the result of an independent random process with a 0.5 probability [7]. Therefore, the unpredictability of the Seed must be ensured in sensitive applications [39]. Consequently, if the generation algorithm and Seed are available, the values produced by a PRNG are entirely predictable. Since in many cases the generation algorithm is accessible, the Seed must be inaccessible and itself be unpredictable from the generated stream sequence [21].

*2.2 Breaking Random Structure*

In pseudo-random generators, an initial seed is used to start the algorithm. This seed is selected randomly. One approach to breaking the structure of pseudo-random generators is determining the initial seed [32]. This is typically because the system must be worked backward through total states to reach the initial starting state. Consequently, another approach is to refer to the algorithm's intermediate states and use them to break the structure. If an algebraic structure exists for generating pseudo-random numbers, by reversing these relationships, the final state leading to number generation can be obtained [22]. Cryptographic attacks can also be used to break random generator algorithm structures; for example, the meet-in-the-middle attack can be used to break the algorithm [40]. In this method, by considering input and output states and an intersection state in the algorithm's middle, we seek a forward and backward collision at the intended intersection state. Overall, these methods aim to reduce examined states and computational complexity. The problem of estimating a pseudo-random generator's structure can be viewed as a cryptography problem with an initial state key. Consequently, block or stream decipher methods can be used to break the structure [32].

*2.3 Statistical and Algebraic Estimation Approaches*

Statistical estimation approaches use statistical tests to reject or accept a tested hypothesis. To examine a selected hypothesis, a statistical test is chosen, and a significance value is used to assess the hypothesis. This approach is statistical and probability-based, not providing a definitive result. The output of statistical approaches is a percentage of match [41]. In algebraic and certain approaches, random numbers are constructed based on a deterministic structure, so the output occurs through breaking algebraic relationships and reversing these relationships [22]. Generally, if an information leak exists in the algorithm, when the state space is very large, a probabilistic approach can be used to avoid examining all states. If deterministic and algebraic relationships exist in part of the algorithm and no information leak is present, using deterministic algorithms is more beneficial.

## 3. Research Method

This section presents analytical discussions and material pertaining to the generation methods and estimation of pseudo-random generators. Additionally, the fundamental structural foundations of PRNGs are introduced and examined.

*3.1 Pseudo-Random Number Generators*

The second type of generator is the PRNG. A PRNG has one or more inputs and generates several pseudo-random numbers [17]. The inputs to a PRNG, known as seed, itself must be random and unpredictable. Therefore, it is preferable to use the output of a RNG as the Seed for a PRNG. Thus, a PRNG can collaborate with an RNG as a companion [42]. The output of a PRNG is typically deterministic functions of the Seed. The term pseudo-random is used due to the deterministic nature of this process [18]. Since each member of a pseudo-random stream can be repeated with its seed, the seed must be stored if reproduction or verification of the pseudo-random stream is needed.

If a pseudo-random stream is correctly constructed, the further we progress in the stream, the more randomness is added to new stream members through repeated transformations [43]. A series of specified transformations can undermine statistical correlations between input and output. Consequently, the output of a PRNG may have better statistical properties and can be generated faster than an RNG. Therefore, using a PRNG instead of an RNG is more operational and optimized in practice, and that is why they are used in many commercial devices.

*3.2 Finite States Machine*

Finite state machines (FSMs) are a computational model derived from the Turing machine but with simpler capabilities [44]. Unlike Turing machines, FSMs have a controller with finite states and a unidirectional moving head for reading/writing on a tape [45]. The input is a bit string, and the output is either the stored bit string or the internal state after reading the input. FSMs are less powerful than Turing machines—for instance, they cannot determine if the number of symbols is prime [44]. However, FSMs are widely used in applications such as pattern matching [46] (e.g., rule-based), modeling logic circuit chains (e.g., Linear Feedback Shift Register), and representing directed graphs [47].

*3.3 Linear Feedback Shift-Register (LFSR)*

LFSR exploits linear functions on previous bits and shifts them to generate new bits. In fact, in the binary field, XOR is the only linear operation on single bits, therefore, its output is predictable. The feedback must be determined in such a way that the output appears pseudo-random. Based on [48], an LFSR is defined using the feedback polynomial $C(X) = 1 - \sum_{i=0}^{L} c_i X^i \in F_q[X]$ where L is called the register length. This register transforms the sequence $(s_0, \dots, s_{L-1}) \in F_q^L$ (called the initial state) into an infinite sequence using the following recursive equation (1):

$$\forall t \geq 0, s_{t+L} = \sum_{i=1}^{L} c_i s_{t+L-i} \tag{1}$$

In this case, the output sequence of an LFSR is $(s_t)_t \geq 0$. The operation of an LFSR is shown in Figure 1.

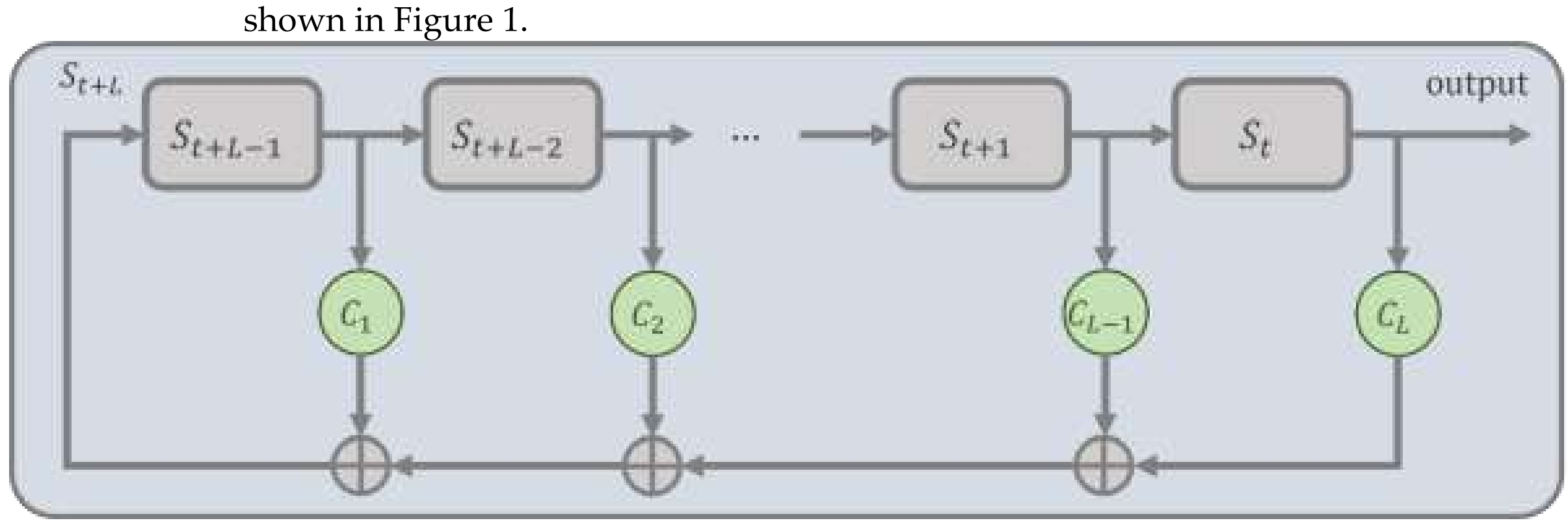

*Figure 1 Structure of LFSR generation process*

The output sequence of an LFSR is uniquely determined by the feedback polynomial coefficients and the initial state. The feedback coefficients $c_1, \dots, c_L$ of an LFSR with length L are specified by the feedback polynomial defined as (2):

$$C(X) = 1 - \sum_{i=1}^{L} c_i X^i \tag{2}$$

These coefficients can also be calculated using the characteristic polynomial in (3):

$$C(X) = X^L C\left(\frac{1}{X}\right) = X^L - \sum_{i=1}^{L} c_i X^{L-i} \tag{3}$$

Once the feedback polynomial is specified and the initial state is estimated, the sequence generated by the LFSR will be predictable. This variant of LFSR, known as Generalized LFSR (GLFSR) [49] generates new bits by utilizing all previous and current states. While this approach significantly extends the period length of the sequence, it necessitates increased memory allocation and requires meticulous seed selection to ensure optimal performance characteristics. In [49], to address seed dependency, twisted LFSR multiplies the current state by a control matrix A, as defined by (4):

$$x_{l+n} \coloneqq x_{l+m} \oplus x_x A, (l = 0,1, \dots) \quad (4)$$

where A is a W×W times matrix (with W based on processing power number of twisted LFSRs), and $x_l$ is a row vector in GF (2) (binary field).

This method increases the period to $2^{nw} - 1$ (compared to $2^n - 1$ f or simple LFSR), reduces dependency on initial state complexity, and allows shorter seeds and less memory usage. It produces more complex outputs, making exhaustive searches harder. Twisted LFSRs are widely used in advanced random number generators like the MT.

### *3.4 MT Generator*

#### 3.4.1 Algorithm Operation

The MT, is a PRNG which is used in generating (pseudo) random numbers. The name "Mersenne Twister" was chosen because its period length is selected as a Mersenne prime number. In mathematics, a Mersenne Prime is a prime number that can be represented in the form $2^n - 1$. Random number generation with MT [43], based on a w-bit word, produces a random number from the set of integers in the range $[0, 2^w - 1]$. Random number generation with MT is based on a matrix with linear recurrence relation on a finite binary field. This algorithm uses a twisted LFSR that operates with state bit representation and matrix rational normal form or Frobenius form. In Frobenius form, a vector space is decomposed into rotational subspaces of matrix A. The main idea is to define a sequence of $x_i$ through a simple recursive relation and then extract numbers in the form $x_i T$. Obviously, T is an invertible matrix in the binary field called the tempering matrix.

*Table 1 Parameters of the MT pseudo-random number algorithm*

| MT parameters | | | |
|---|---|---|---|
| **Parameter** | definition | MT Version | |
| | | MT 19937 - 64 | MT 19937 - 32 |
| **w** | word size in bits | 64 | 32 |
| **n** | degree of recurrence | 312 | 624 |
| **m** | middle word | 156 | 397 |
| **r** | separation point | 31 | 31 |
| **a** | matrix coefficient | 0xB5026F5AA96629E9 | 0x9908B0DF |
| **u** | tempering right shift | 29 | 11 |
| **d** | tempering mask | 0x5555555555555555 | 0xFFFFFFFF |
| **s** | tempering left shift | 17 | 7 |
| **b** | tempering mask | 0x71D67FFFEDA60000 | 0x9D2C5680 |
| **t** | tempering left shift | 37 | 15 |
| **c** | tempering mask | 0x FFF7EEE000000000 | 0xEFC60000 |
| **l** | tempering right shift | 43 | 18 |

```
Mersenne Twister algorithm generation operation

Initialize constants (w,r,f,u,d,s,b,t,c) based on Table (2)
Define matrix A
Initialize array of n w-bit values using w-bit seed as x[0]
x[0] ← seed
Loop i from 1 to n-1
    x[i] ← f * (x[i-1] ^ (x[i-1] >> (w-2))) + i
Initialize sequence x as an array of w-bit values (internal state)
Loop for k from 0 to n-1:
      xk^u ← xk // w-r
      xk+1^l ← xk+1 % r
      xk+n ← xk+m ^ (xk^u||xk+1^l) A
      y ← xk+n ^ ((xk+n >> u) & d)
      y ← y ^ ((y << s) & b)
      y ← y ^ ((y << t) & c)
      z ← y ^ (y >> 1)
      until s + t ≥ ⌊w/2⌋ - 1
```

*Figure 2 MT generation algorithm pseudo code*

Based on Table 1, parameters of MT random number generation algorithm were set. So, $2^{nw-r}-1$ is a Mersenne prime. Based on [43], the structure needed for implementing MT is an array of n values, each w bits long. For initial array assignment, a w-bit seed value is used to determine $x_0$ to $x_{n-1}$. The value $x_0$ is the seed, and subsequent random sequence values are obtained as (5):

$$x_i = f \times \left(x_{i-1} \oplus \left(x_{i-1} \gg (w-2)\right)\right) + i \tag{5}$$

The value i changes from 1 to n-1. Thus, the first random number generated by the algorithm depends on $x_n$ The constant value f also depends on the generator, so it's not seen in the algorithm. In MT19937-32 and MT19937-64, their value is, respectively, 1812433253, and 6364136223846793005. Then, the sequence of x's is defined as a set of w-bit values with the recursive equation (6):

$$x_{k+n} := x_{k+m} \oplus \left(\left(x_k^u || x_{k+1}^l\right) A\right), (k = 0,1, \dots) \tag{6}$$

Where || indicates bit vector concatenation (with upper bits on the left), $\oplus$ is the bitwise XOR operator, $x_k^u$ means the upper w-r bits of $x_k$, and $x_{k+1}^l$ is the lower r bits for $x_{k+1}$.

In this case, the twist transformation of matrix A to normal logical form will be as (7):

$$A = \begin{pmatrix} 0 & I_{w-1} \\ a_{w-1} & (a_{w-2}, \dots, a_0) \end{pmatrix}. \tag{7}$$

In the above relation, $I_{w-1}$ is an identity matrix (square with w-1 rows and columns). Using the rational normal form for matrix A has the advantage that multiplication can be shown as (8):

$$xA = \begin{cases} x \gg 1 & x_0 = 0 \\ (x \gg 1) \oplus a & x_0 = 1 \end{cases} \tag{8}$$

Here, the XOR bit operator is used as addition. The MT algorithm employs a compensatory mechanism to address dimensional reduction in uniform distribution through sequential transformation steps. This process is mathematically equivalent to implementing a similarity transformation on matrix A, expressed as $A = T^{-1}AT$, where T represents an invertible transformation matrix. This approach enables the algorithm to maintain distributional properties despite the inherent constraints of computational random number generation. In this algorithm, the tempering transformation is written as (9):

$$\begin{aligned} y &:= x \oplus \big((x \gg u) \& d\big) \\ y &:= y \oplus \big((y \ll s) \& b\big) \\ y &:= y \oplus (((y \ll t) \& c)\ ) \\ z &:= y \oplus (y \gg l) \end{aligned} \tag{9}$$

Where x is the next value in the number sequence, y is a temporary initial value, z is the calculated value from the algorithm, and the symbols ≪ and ≫ represent left and right bit shifts. The & symbol is the bitwise AND operator. Based on the properties in twisted LFSR, condition (10) provides uniform distribution for upper bits:

$$s + t \geq \lfloor \frac{w}{2} \rfloor - 1 \tag{10}$$

Pseudo-code presented in *Figure 2*, There are other methods for generating the final random number besides the 32-bit and 64-bit methods. In some versions, the final random number is 31-bit (actually, one bit is removed in binary space). For example, newer versions of MATLAB software use the MT19937ar version. In this version, the output is produced with 53-bit resolution as (11):

$$x = (a * 67108864 + b)/(9007199254740992) \tag{11}$$

Where a and b are 32-bit random numbers from the introduced MT method that have been reduced to 27 and 26 bits, respectively. With the discarding of bits from a and b, estimation becomes harder and needs to further work.

*3.5 Combining pseudo- random number streams (LFSR and MT)*

To enhance sequence complexity, different PRNG methods—such as LFSR and MT—can be combined to leverage their strengths. This combination can lead to either linear or nonlinear increases in complexity. While greater complexity makes it more challenging to hack or predict the generated streams, the quality of randomness in the output, i.e., generated output pseudo-random stream through combination of different PRNGs, remains largely unaffected in practice. In some cases, although complexity increases, the randomness actually decreases.

Note that there are multiple ways to combine pseudo-random number streams to each other, but these combinations should hold some condition to enhancing randomness and complexity (linear or non-linear). Common combination techniques are as follows:

3.5.1 Linear Combination (XORing Outputs): This is the simplest approach, where the bit streams from multiple LFSR generators are combined using the XOR (exclusive OR) operation. This method adds the outputs of the LFSRs together in a binary field, essentially performing modulo-2 addition on each bit. This method is easy to implement but might not significantly increase the complexity if the individual LFSR generators have predictable structures or short periods. In this case, the resulting combined sequence can be easily broken down.

3.5.2 Non-Linear Combination: Non-linear combination techniques aim to introduce more complexity and randomness by using non-linear elements in the combination process. This can make the resulting sequence harder to predict compared to simple XOR combinations.

**a) Multiplication in Galois Field:**

This method involves multiplying the outputs of LFSRs within a Galois Field (GF(q)), a finite field with a specific number of elements. The complexity of the resulting sequence depends on the degrees of the polynomials associated with each LFSR and their common factors. This method has potential for generating highly complex sequences but can decrease randomness if the multiplication results in a stream of zeros. The selection of appropriate polynomials and understanding the properties of Galois Fields is crucial for effective implementation [50].

**b) Flip-Flop Combination:** In [51], the outputs of two LFSRs are fed into a flip-flop, like a JK flip-flop. The previous output bit is used to control the flip-flop, introducing non-linearity into the combined sequence. This method Adds non-linearity, potentially increasing period and complexity but can increase correlation between the combined LFSR outputs, reducing overall randomness. Careful design is needed to avoid predictable patterns and ensure sufficient randomness.

3.5.3 Clock Control: one LFSR's output can control the clock of another LFSR [52]. This method introduces non-linearity and can increase complexity but requires careful design to avoid predictable patterns in the clock control mechanism. If the clock control becomes predictable, it can undermine the randomness of the combined sequence[39].

When designing pseudo-random sequence generators, careful consideration must be given to combination methodologies. Introducing non-linear elements can enhance unpredictability, but poorly implemented combinations risk reducing randomness and increasing predictability. The choice of method—whether linear (e.g., XOR) or non-linear —should align with the desired balance of complexity, randomness, and application-specific requirements. Success hinges on a rigorous understanding of the underlying mathematical principles to optimize the trade-offs between security and efficiency.

In practical LFSR implementations, developers often prioritize processing speed and reproducibility by using shorter initial seeds and polynomials. While this simplifies bit sequence regeneration, it inherently weakens cryptographic resilience, rendering the structure susceptible to predictive attacks and cryptanalytic exploitation. Similarly, other pseudo-random generators like the MT lack inherent randomness; their perceived unpredictability relies solely on the computational difficulty of reverse-engineering their deterministic patterns.

Quantum-generated sequences produced via QRNGs inherently preserve their randomness across all temporal windows and sequence lengths, as their unpredictability originates from fundamental quantum properties rather than algorithmic complexity [39]. QRNGs leverage quantum effects—such as quantum superposition, uncertainty, fluctuations, and spontaneous emission, which arise from the quantum vacuum—to ensure true inherent quantum mechanically randomness. Even when a transient pattern is artificially introduced within a limited segment of a quantum sequence, such patterns fade over subsequent windows, rendering them non-repeatable and cryptanalytically insignificant [39]. This fundamental distinction highlights the crucial advantage of QRNGs in security-sensitive applications, where conventional pseudo-random generators—despite their efficiency and speed—remain susceptible to structural vulnerabilities. Recent breakthroughs in high-speed QRNG technology, capable of achieving gigabit-per-second data rates, have facilitated their practical implementation across various domains. QRNGs are now deployed in cryptographic security, secure communications, microwave noise-based sensing, and real-time data encryption, providing a level of randomness that is inherently resistant to external manipulations. Unlike PRNGs, QRNGs derive randomness from genuine quantum phenomena rather than deterministic algorithms, making them immune to cryptanalytic attacks and the predictive capabilities of emerging quantum computers. Their scalability and robustness have positioned them as an integral component of next-generation security frameworks, including quantum-safe encryption, quantum networking, and advanced quantum sensors. As continuous innovations in hardware and optimization enhance QRNG efficiency, their adoption is rapidly expanding within consumer electronics, financial systems, and industrial applications.

On the other hand, with the ongoing advancements in quantum computing, traditional cryptographic methods that rely on PRNGs face increasing risks. Quantum algorithms, such as Shor's and Grover's, have demonstrated the potential to compromise encryption schemes dependent on PRNG-derived keys. This growing threat underscores the urgency of transitioning from classical RNGs and PRNGs to QRNG-based security solutions. Thus, by harnessing intrinsic quantum randomness, QRNGs establish a more resilient foundation for cryptographic security, ensuring that encryption protocols remain impervious to quantum-driven attacks. The accelerating integration of QRNGs into global security infrastructures will play a pivotal role in safeguarding digital assets, secure communications, and financial transactions against future cryptographic vulnerabilities in an era dominated by quantum computing.

### *3.6 Prediction Procedure of PRNGs or RNGs*

PRNGs utilize deterministic algorithms, whereby systematic exploration of the parameter space can reveal the future sequence of the stream. However, the computational complexity required to determine these sequences varies significantly across different implementations. LFSR and MT streams are constructed upon specific algorithms and algebraic structures, rendering them inherently predictable. Consequently, future values within these streams may be compromised, effectively negating the stream's randomness properties, and constituting a successful cryptanalytic attack.

Methodologies developed for predicting pseudo-random streams can be categorized into two principal approaches: pattern-based searching methods (characterized by deterministic precision) and correlation-based methods (characterized by probabilistic precision). Pattern-based searching methodologies yield binary outcomes - either complete prediction accuracy or total failure. In contrast, correlation-based methodologies provide probabilistic predictions within specified confidence intervals, quantifying the likelihood of future values falling within particular ranges.

#### 3.6.1 Prediction Method of LFSR

In practical communication systems, output data (e.g., text or encoded information. For example, the output of the LFSR serves as the input to the prediction algorithm) frequently exhibits statistical bias, characterized by non-uniform probability distributions of binary digits. For instance, certain bits may demonstrate higher probabilities of being 0 or 1 due to inherent language statistics or specific encoding format requirements (e.g., ASCII). This phenomenon primarily stems from the fact that the output data bits are not uniformly random. Consequently, in the case of LFSR-generated streams, true randomness is not achieved, resulting in biased output sequences. These conditions also apply to the input parameters. The inputs of practical communication systems (particularly LFSR-based systems) exhibit statistical biased properties like those observed in the outputs of these system. The objective is to reconstruct the feedback polynomial of the LFSR, given only the biased input and the accessible observed LFSR output sequence.

Now, let $(x_t)_{t\geq 0}$ denote the input sequence to the LFSR, which is unknown (or known and biased) but satisfies $pr[x_t = 0] = \frac{1}{2} + \epsilon$, where $\epsilon \neq 0$ represents the bias. The output sequence $(y_t)_{t\geq 0}$ of the LFSR is observed, and the task is to reconstruct the feedback polynomial of the LFSR. The reconstruction relies on a statistical test that exploits the bias in the input sequence to detect the feedback polynomial of the LFSR [53]. The key idea is to compute a random variable Z based on the output sequence $(y_t)_{t\geq 0}$ and check if it follows a Gaussian distribution with specific mean and variance.

Based on [39], [53], the theoretical foundation for the reconstruction method presented. Let the feedback polynomial of the LFSR be $1 + \sum_{j=1}^{d-1} X^{i_j}$, where $i_1 < \cdots < i_{d-1} = L$. Define the random variable $z_t = y_t \oplus \left( \bigoplus_{j=1}^{d-1} y_{t-i_j} \right)$ for $t \geq L$. Subsequently, the random variable $Z = \sum_{t=L}^{N-1} (-1)^{z_t}$ follows a Gaussian distribution characterized by $(N - L)(2\varepsilon)$ and $(N - L)(1 - (2\varepsilon)^2)$, representing the mean and variance of the distribution, respectively. Algorithm's Pseudo code presented in Figure 3. Also, diagram of algorithm is shown in Figure 4. This result is derived from the fact that $z_t$corresponds to the input bit $x_t$,which is biased. The Gaussian distribution of Z allows us to perform a statistical test to determine whether a candidate polynomial is a valid feedback polynomial. The reconstruction algorithm involves testing all possible feedback polynomials of degree up to $L_{max}$, where $L_{max}$ is the maximum assumed length of the LFSR that be chosen by user. For each candidate polynomial, the algorithm computes the value of Z and compares it to a threshold T to determine if the polynomial is a valid feedback polynomial. In algorithm, for each candidate polynomial of degree up to $L_{max}$, compute Z and calculate maximum of the this is the polynomial that is considered as a valid feedback polynomial. So, the first polynomial that satisfies the condition is returned as the feedback polynomial of the LFSR. Sparse polynomials (e.g., trinomials or binomials) are easier to recover, and the algorithm is efficient in such cases. For example, in practical systems like synchronous optical networking (SONET), where the feedback polynomial is $1 + X^{43}$, or ITU V.34 (a modem standard that enables data transmission speeds of up to 33.6 Kbps over phone lines), where the polynomial is $1 + X^{18} + X^3$, the algorithm performs well.

However, the complexity increases significantly for non-sparse polynomials, as all possible feedback polynomials must be tested. This makes the algorithm less efficient for structures with more coefficient feedback polynomials. The used method requires testing all possible feedback polynomials, which increases the complexity, especially for polynomials with much coefficient. However, the used method requires fewer output bits compared to other cases because the statistical bias is more pronounced.

**Prediction Method of LFSR**

```
Input LFSR, Deg_Max, Term_min, Term_max
If LFSR length < 2 * Deg_Max, then raise an error
If Term_max < 2, then raise an error
If Term_min < 2, then raise an error
Initialize T_Z∈R³ (T_Z=0)
Initialize t = 1
Loop for d from Term_min to Term_max
    D ← Choose d-1 from 1:Deg_Max
    Loop for i from 1 to size(D)
        z ← LFSR
        Loop for j from 1 to d-1
            z ← z + circshift(LFSR,-D_{i,j})
        z ← z(D_{i,j}+1:end) % 2
        Z ← Σ_{i=1}^{length(z)} (−1)^{z_i}
        T_Z_t ← [d,i,Z]
        t ← t+1
I ⟵ Find the maximum indices of normalized T_Z
Poly ← Extract polynomial coefficients and set the Poly output
Init ← flip(LFSR)
LFSR_New ⟵ Generate a new LFSR sequence using Poly and Init
Compute match-points and adjust match-point based on conditions
Return Poly, Init, and match-point(MP)
```

*Figure 3 LFSR structure breaking algorithm pseudo code*

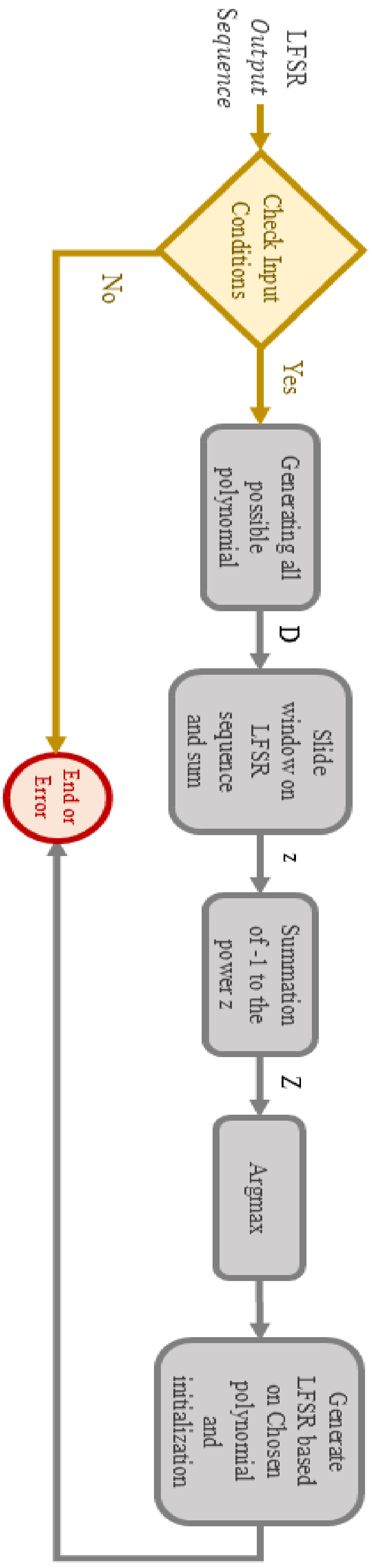

*Figure 4 Block diagram of the method used for breaking the LFSR structure*

*3.6.2 Prediction Method of MT*

**Deterministic Prediction of Mersenne-Twister**

```
Giving y as input to the algorithm
if length(y) < 20160
      throw an error
Initialize i ← 0 ,  Initialize input_to_model_stream ← []
Loop until 625*32 + i < length(y)
      z← Extract i to i+624*32 bits from y
      z1 ← bit2int(z , 0) # Convert z to 32-bit integers
      z1 ← untempering(z1)
      MT ← Generating a pseudo-random number and checking
      whether the generated number is equal to
      i + 624 * 32: i + 625 * 32 or not.
      If MT is not empty
            MSB ← 0, Offsetbit ← i
            Return
      z2 ← bit2int(z , 1) # Convert z to 32-bit integers
      z2 ← untempering(z2)
      MT ← Generating a pseudo-random number and checking
      whether the generated number is equal to
      i + 624 * 32: i + 625 * 32 or not.
      If MT is not empty
            MSB ← 1, Offsetbit ← i
            return
      i ← i+1
If MT is empty
      MT← -1, Offsetbit ← -1, MSB ← -1
      return
```

*Figure 5 MT structure breaking algorithm pseudo code*

Twister algorithm consists of three stages: initialization, twisting, and tempering. Estimating future terms requires implementing reversible steps:

To predict the continuation of the MT sequence, an estimation of the seed is assumed. To estimate the initial seed, it is also necessary to obtain 624 internal 32-bit states. The internal states of the MT algorithm can be obtained using algebraic and bidirectional equations, but more operations are needed to estimate the seed. However, discovering the 624 internal states is sufficient to estimate the next generated bit sequence.

According to [43], to generate a random number using this method, all 624 internal 32-bit states are first filled, and then, to generate subsequent numbers, these states are modified one by one in sequence. As a result, by accessing these states, the rest of the random sequence can be generated based on clocking the LFSR. Therefore, in [54], the need to estimate the seed is eliminated, and by discovering the intermediate states, the output can be generated according to the set seed. However, to estimate the next bit of the sequence, it is necessary to know all the intermediate states (at least 624*32 bits).

So, for predicting the next bit of MT pseudo random sequence generator, output sequence is in access. By inversion of the tempering, which is the so-called untempering, the generated sequence transformed to twisted states. In this stage, by clocking the bits, pseudo random numbers were generated one-by-one. For inversion, all tempering actions were inverted algebraically. At this time, testing the generation of 625th state, by the last 624th states, could determine whether the pattern of MT algorithm was existed or not. Thus, reversed states feed to a MT generation algorithm for generating next outputs. If matching done so the MT structure was predicted. The prediction success could be evaluated as a percentage, not as a binary outcome. This is meaning that for example algorithm succeeded to predicting next bits in 60%, so match-point rate in this case is 60%. The MT prediction algorithm's Pseudo code and block diagram are respectively presented in Figure 5 and Figure 6.

One of other considered scenarios is flip the sequence for feeding to algorithm. these cases must include in checking states. So, for exact checking, sequence and its flipped version are searched for finding the algorithm.

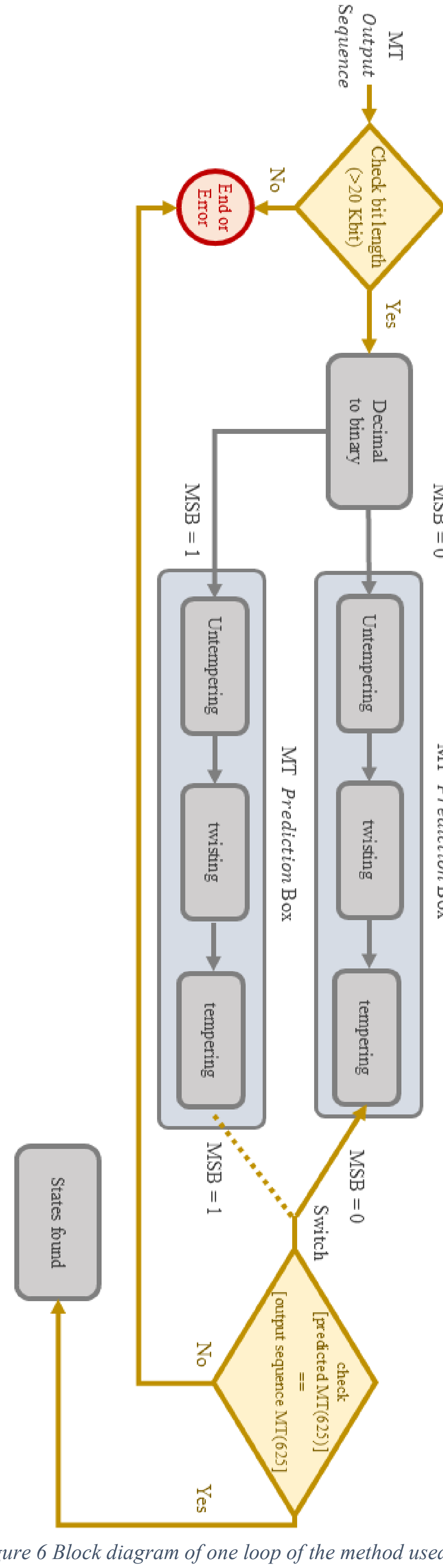

*Figure 6 Block diagram of one loop of the method used for breaking the MT structure*

3.6.3 Prediction of Combined Multiple LFSR or LFSR-MT

Linear complexity for prediction of next bits of pseudo random number sequence increases as linear combination used for output of some LFSR. The concatenation of pseudo-random sequences represents one methodological approach to combination techniques. Mathematically, the juxtaposition of two-bit sequences is equivalent to shifting the second sequence by the length of the first sequence, followed by the XORing of these two-bit sequences. Therefore, the concatenation of two-bit sequences can be represented by the equation (12):

$$\begin{cases} a_{n+m} = a_n || 0_m \\ b_{n+m} = 0_n || b_m \end{cases} \rightarrow a_{n+m} \oplus b_{n+m} = a_n || b_m = a_n * 2^m + b_m \tag{12}$$

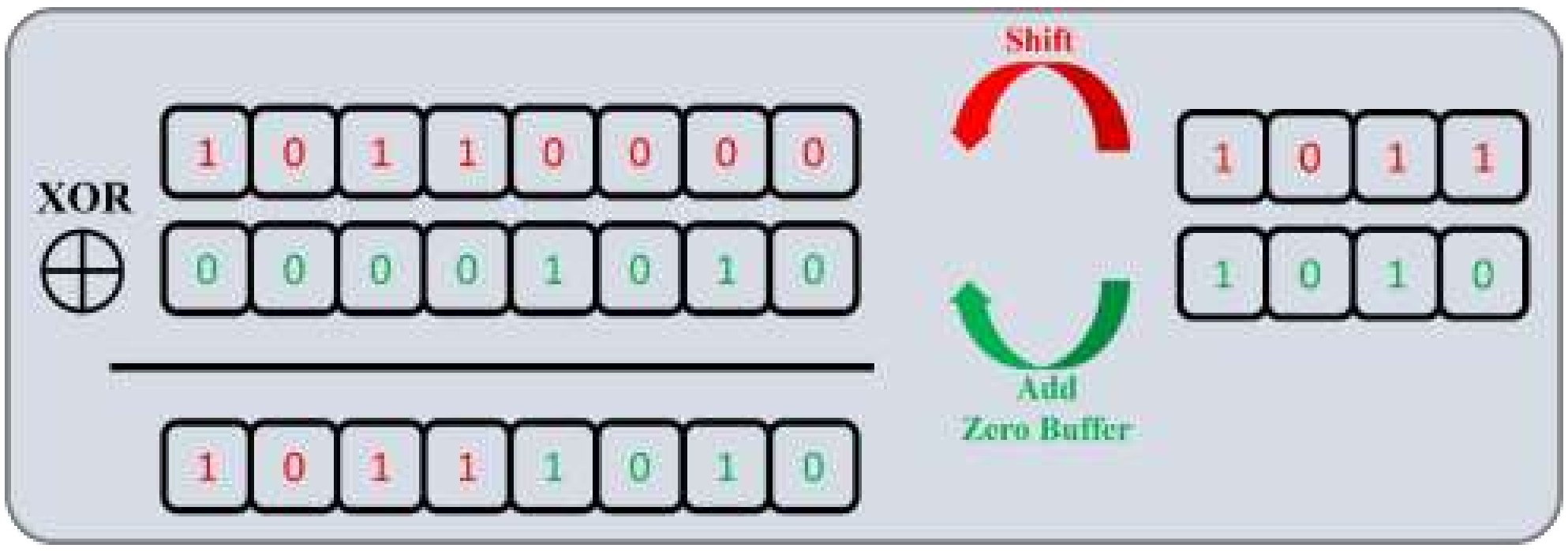

*Figure 7 juxtaposition of two sequence of bits*

Combining two pseudo-random bit sequences—by shifting one sequence by the length of the other and XORing them— (Figure 7) is mathematically equivalent to increasing their vector dimensions and performing summation. While this combined sequence cannot be generated by a single LFSR (with primal initial states), the operation itself constitutes a linear combination within the LFSR framework (over GF (2)). Shifting the first sequence effectively expands the second sequence's dimension, and XORing them integrates their outputs into a composite linear structure. This principle generalizes to concatenating n-LFSR outputs or pseudo-random sequences from any platform, linearly amplifying computational complexity. However, such linear enhancements remain cryptographically vulnerable, as the computational effort required to compromise the output retains a linear order of complexity [39] .

To manage multi-LFSR combinations, precise boundary detection between individual generator outputs is critical. A "match-point rate" metric evaluates polynomial fit by analyzing correlation within a sliding window that iterates across the bit stream. The optimal boundary is identified when this rate peaks before degradation begins. Accurate boundary partitioning is essential; errors here propagate into incorrect predictions, underscoring the need for robust detection techniques.

Ultimately, while combining multiple LFSRs or pseudo-random generators linearly escalates complexity, it fails to transcend linear-order vulnerability. Introducing non-linear operators (e.g., Galois field multiplication or flip-flop logic) can enhance unpredictability, but excessive non-linear combinations risk amplifying correlations between components and outputs. This creates a trade-off: non-linear complexity improves security but demands careful design to avoid unintended correlations that undermine randomness. Thus, effective hybrid architectures must balance non-linear enhancements with controlled correlation to achieve resilience.

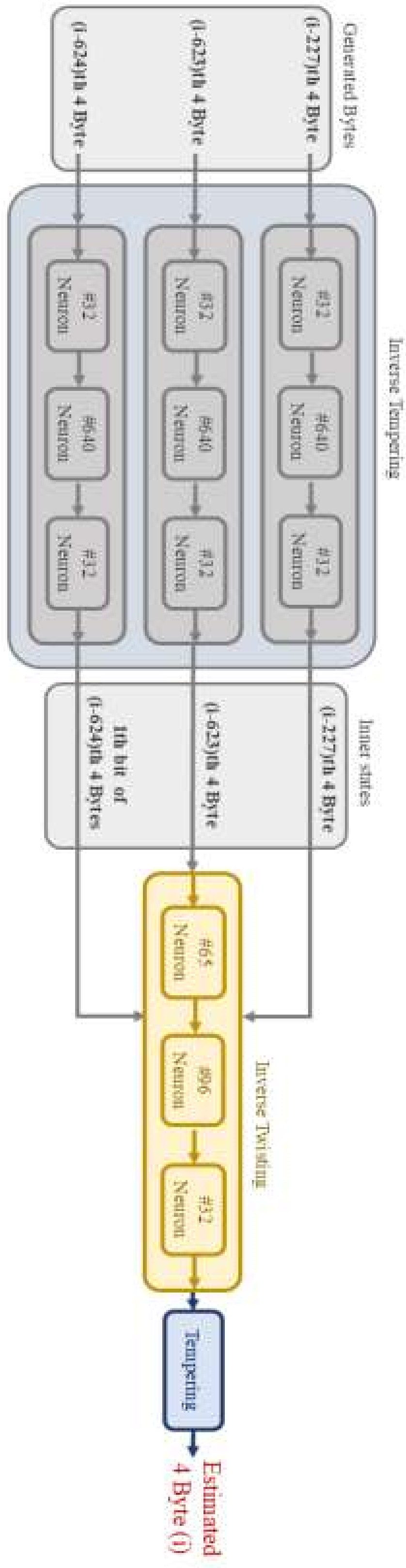

*Figure 8 Schematic of deep learning diagram*

3.6.4 Deep Learning-Based Prediction of MT

Deep learning (DL) networks demonstrate significant potential for predicting pseudo-random stream structures through their robust feature extraction capabilities. These networks can effectively identify various types of correlations, thereby enabling bit-to-bit relationship mapping without requiring prior information about the underlying structure. Implementation necessitates training a DL model on a generated dataset specifically designed for this purpose [55]. The model architecture typically accepts a fixed-length window of previous MT outputs as input and generates predictions for subsequent output bits based on learned patterns within the temporal sequence.

To predict the pseudo-random sequences generated by the MT without using the seed, access to 624 output states is required [43]. The process of generating pseudo-random numbers in the MT is deterministic, so using machine learning-based methods does not probably reduce the complexity of prediction. The limitation in predicting the MT sequence lies in the minimum length of the generated sequence required for prediction. To investigate the minimum length needed for predicting the MT sequence, non-deterministic methods can be employed. Artificial intelligence-based methods can reduce the complexity of search in unknown spaces by discovering correlation and new pattern [55].

In [34], a neural network model is trained for prediction (Figure 8). This model consists of 4 separate neural networks. Overall, processing states i, i-624, i-623, and i-227 leads to the prediction of new states, thereby generating new random numbers. Therefore, if the goal is to reduce the minimum required length for prediction, the dependency on states i-624 and i-623 must be eliminated. In this case, the minimum required length for prediction could reduce to 227*32 bits.

The question that arises here is whether statistics methods (such as machine learning based methods) can extract new relationships for better prediction. In the MT method, for generating the 625th random number or state, only states i-624, i-623, and i-227 are involved, and no other states from the initial 624 states contribute to the generation of new number. Therefore, the minimum number of states involved in predicting the sequence is these three states, and there are no other unique correlations to discover new states. Thus, the deep learning-based method used here also processes these three states. Consequently, the use of deep learning in this context is merely training a model to fit the deterministic prediction function provided in the second phase. As a result, the minimum length required for estimating the MT method is 624*32 bits, which is also used in this method. Reducing the minimum required length is a separate research topic that can be explored independently using innovative reducing complexity methods.

A notable point in this section is the degree of correlation between the output of the proposed method in [34] and the three states i-624, i-623, and i-227. Based on the analysis in [34],the random number generated by the MT method has little correlation with the first 31 bits (Least Significant Bit) of state i-624. This conclusion is derived from evaluating and monitoring the weights in the trained neural network. Therefore, the input to the second part of the network is only 65 bits.

The model consists of 4 parts, three of which map the input bits i-624, i-623, and i-227 to the states before the tempering phase, and one part derives the i-th state from them before twisting, which is 65 bits (32 + 32 + 1). In the proposed method, if a sequence of 624 32-bit numbers is considered, the random numbers i-624, i-623, and i-227 are fed as input to the model, and the 625th number is generated. Finally, to generate a random number, the obtained state must be tempered to produce the (pseudo-)random output.

In training the first three parts, which are similar, the generated random number is used as training data, and its corresponding state before tempering is used as the label. In training the second part, states i-624, i-623, and i-227 are used as training input (32bit number), and state i is used as the label (though, as mentioned, only the 1 MSB of state i-624 is used).

In the model, for both parts of the network, two linear layers with ReLU and Sigmoid activation functions are used, respectively. The training data accuracy and correctness in the training and testing process have reached 100%. This means the neural function used perfectly fits with the algebraic function used. In conclusion, using a neural network fitted with the algebraic function does not provide a speed advantage here.

## 4. Results

*4.1 Software window*

The software we designed has two main modes: one for generating pseudo-random sequences and another for breaking down and hacking their structure.

4.1.1 Random Sequences Generation Mode

This mode can generate pseudo-random bit streams using LFSR, Mersenne Twister 19937, and Mersenne Twister Residual 53 structures.

For LFSR-based streams, you can define the generating polynomial and the initial state (Figure 9). After setting the generator's structure and initial state, the software produces pseudo-random bits based on the length you've specified. Similarly, for Mersenne Twister-based streams, you can observe the pseudo-random output by defining an initial seed (Figure 10).

The generation mode also includes a section for creating combined pseudo-random numbers by mixing several LFSR structures of different lengths. This feature is useful for testing attacks on linear combinations of multiple LFSRs. In this mode, you can also combine several LFSR and Mersenne Twister generators to create a hybrid pseudo-random stream. As mentioned in section 3.4, combining pseudo-random streams from different origins can increase linear complexity.

4.1.2 Hacking Mode

In this mode, you can input data from several sources: a generator, a text file, a .mat file, or direct definition. This input must be a bit string consisting of only 0s and 1s.

We have developed several algorithms for hacking the polynomials of LFSR and the 19937 version of Mersenne Twister. To estimate the LFSR polynomial, you can set the maximum polynomial degree and the maximum and minimum number of polynomial terms to narrow down the search space. This mode also includes a feature to break the structure of the Mersenne Twister pseudo-random generator without needing to estimate the initial seed.

For hacking combined pseudo-random streams from LFSR and Mersenne Twister, the implemented algorithm also requires you to define a search space for finding the LFSR polynomial. Therefore, you must also specify the maximum search degree and the maximum and minimum number of terms in this mode.

4.1.3 Outputs and Analysis

The outputs from both the generator and hacking modes can be saved as a text file. In the hacking mode, three windows display the hackable random string, the estimated bit string for that sequence, and the bitwise XOR difference between the two.

The estimated polynomial and initial states for LFSR structure estimation, a success message for a successful hack, a message indicating no pattern was found, the time spent for each algorithm, and necessary warnings for each estimation mode are all shown in a text display on the right side of the software.

The LFSR polynomial estimation algorithm is based on a statistical method, so its output is probabilistic. A criterion called "match-point" is used to measure how well the generated and estimated bit strings match. This criterion represents the output of a statistical experiment, so the result is a statistic. However, the output of the Mersenne Twister estimation algorithm is a search for a specific pattern, and its output message is either "100% success" or "0% failure" (pattern not found).

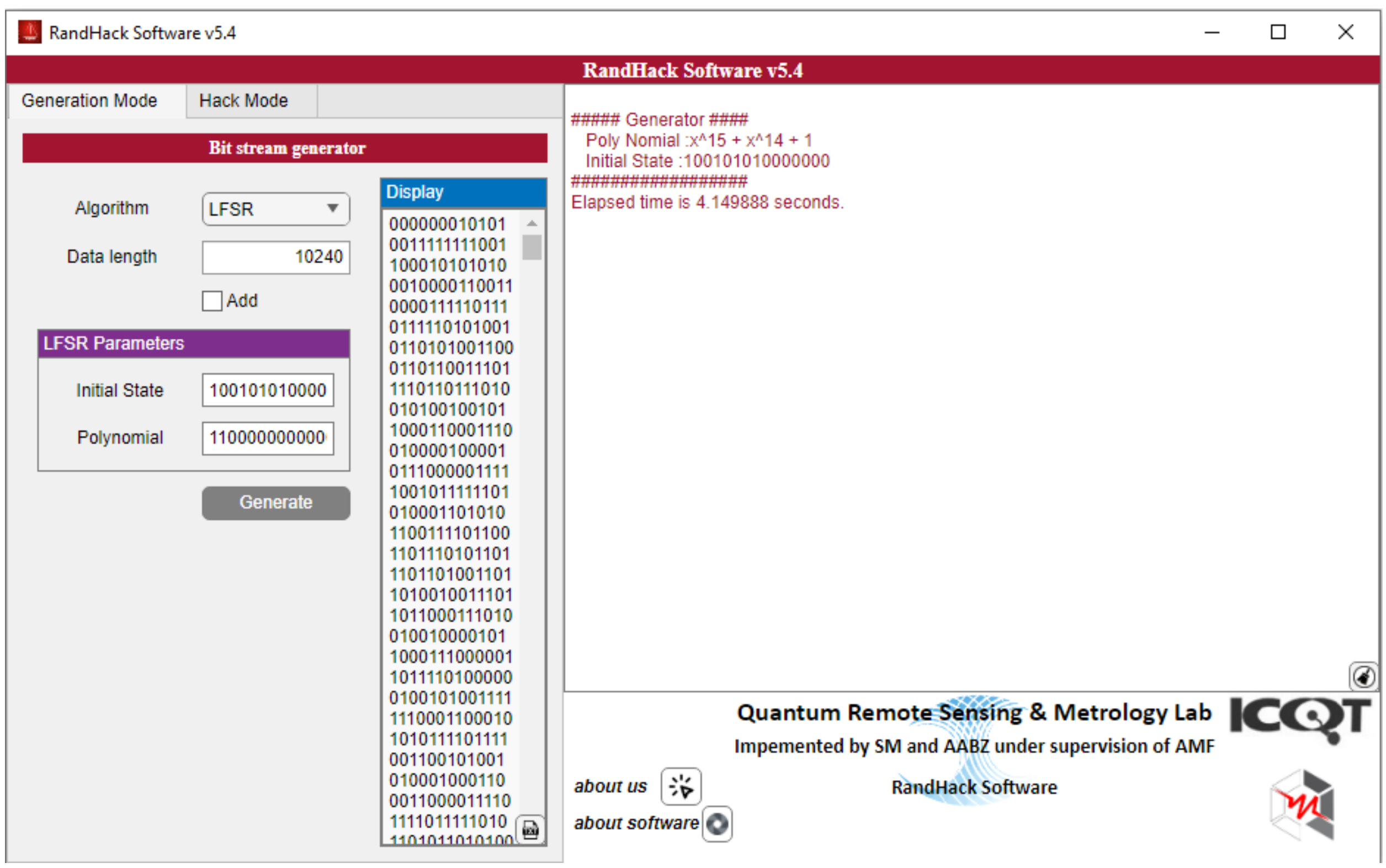

*Figure 9 LFSR stream Generation mode*

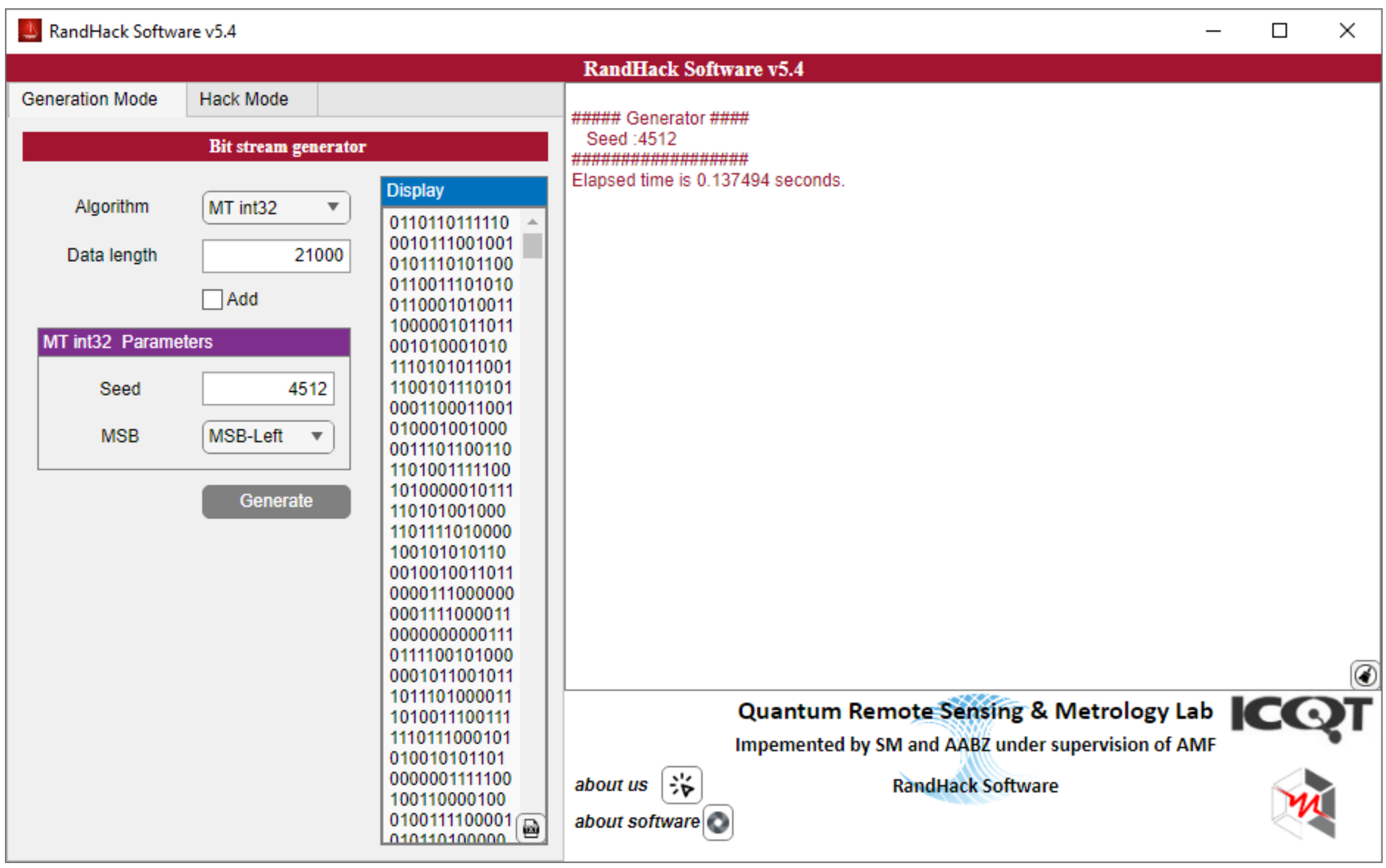

*Figure 10 MT stream Generation mode*

*4.2 Performance*

4.2.1 Deterministic discovering algorithm

Note that in pseudo-random number generation methods based on deterministic algorithms, precise and deterministic relationships exist between different components constructing the algorithm. Therefore, a deterministic method for discovering patterns is used, which can be expressed through algebraic equations. These algebraic equations are also reversible, so their algebraic inverses exist and can be retrieved through reversal. In methods based on deterministic structure search, due to the existence of algebraic relationships, precise pattern exploration occurs in the output bit stream, as the goal is to discover an exact relationship in the bit stream. If this relationship does not exist, the desired pattern remains undiscovered, and the generated bits do not correspond to the searched method.

Therefore, in deterministic generation, a specific pattern exists in the generated sequence. In the MT sequence estimation method, the search is conducted to discover the MT generator pattern. According to results presented in [34], three state vector conditions (i-624, i-623, and i-227) are sufficient to construct the next state and continue the bit stream. Consequently, by examining these three states, new inner states (i) can be predicted by the developed software as can be seen in Figure 11 ,Figure 12 ,Figure 13. As a result, by analyzing the relationships between these three consecutive states in the bit stream, future outputs can be anticipated.

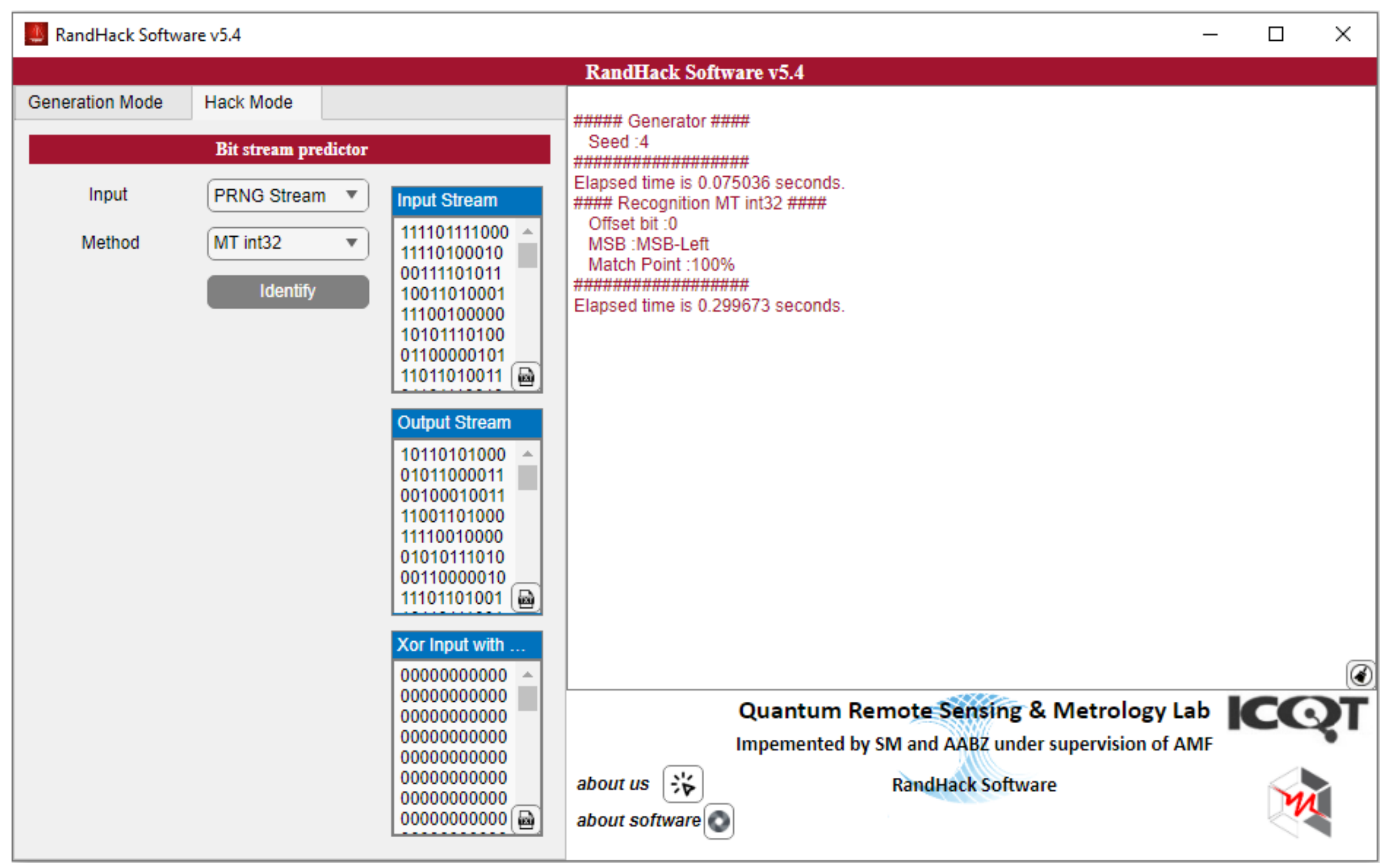

*Figure 11 Breaking the structure of the MT pseudo-random sequence in the designed software with seed 4*

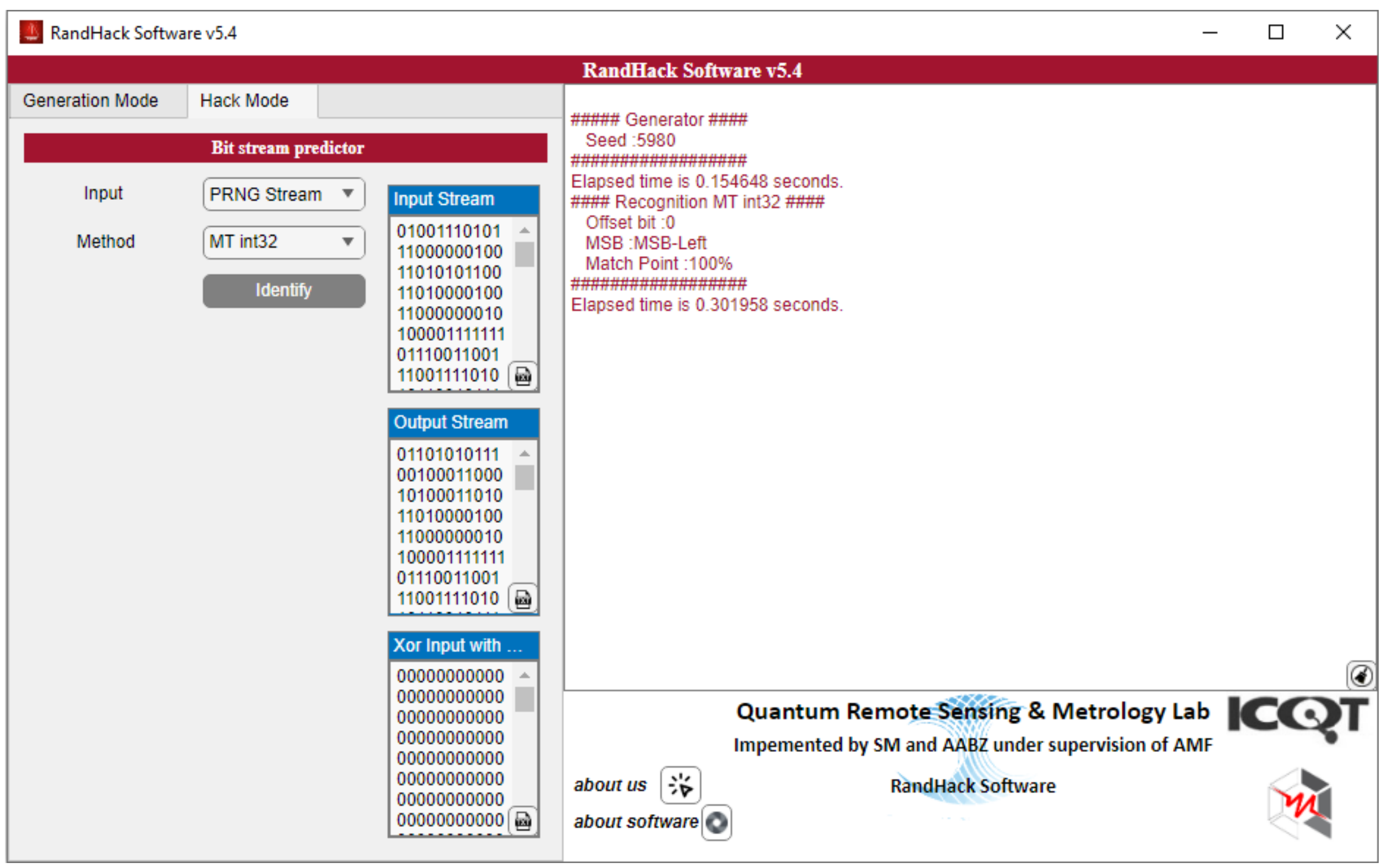

*Figure 12 Breaking the structure of the MT pseudo-random sequence in the designed software with seed 5980*

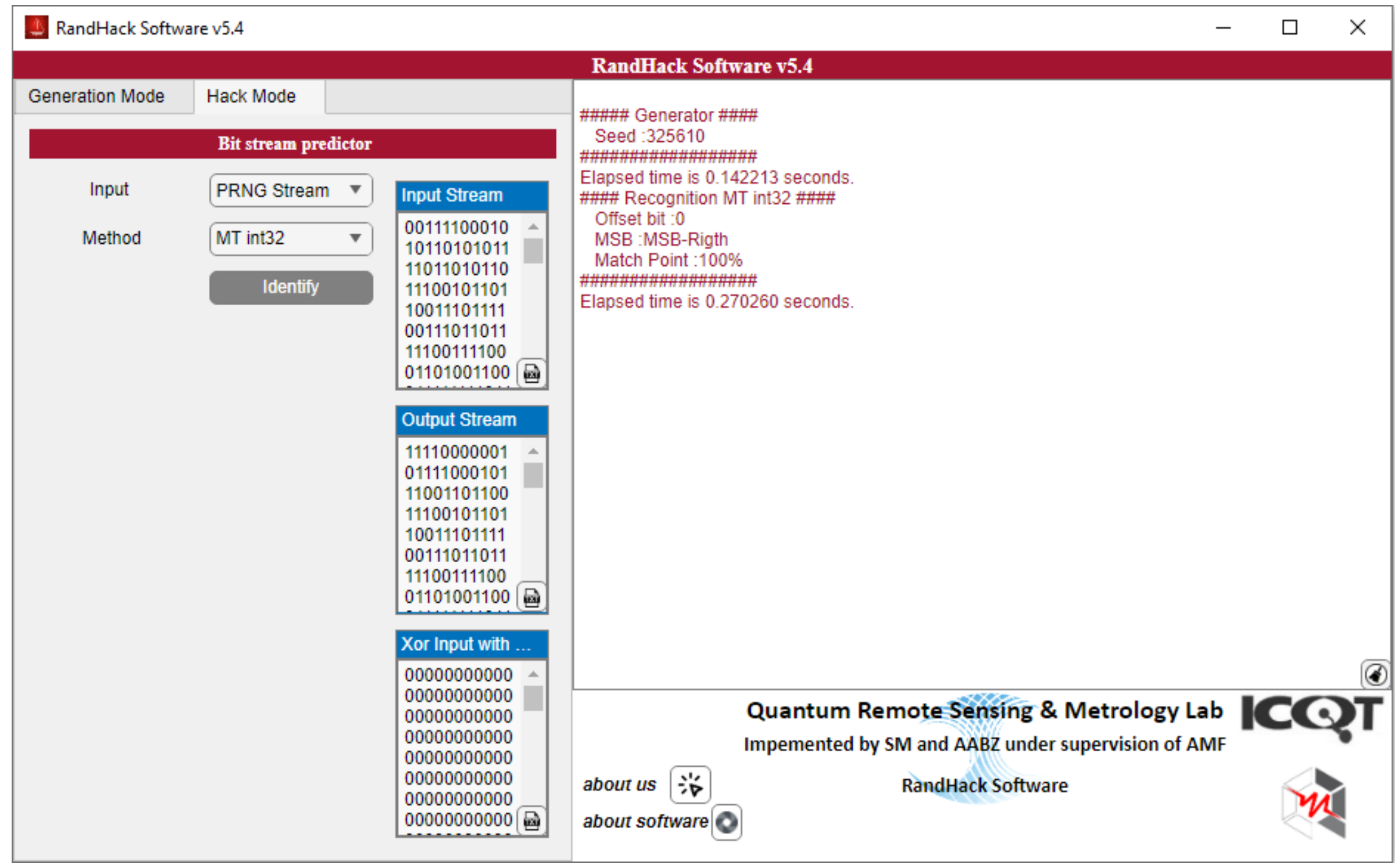

*Figure 13 Breaking the structure of the MT pseudo-random sequence in the designed software with seed 325610 and MSB-Right*

4.2.2 Statistics-based Prediction

The method for estimating the initial value and characteristic polynomial of LFSR is based on a statistical test. Our developed software uses a sliding window on the available bit stream in the applied method. Consequently, the computational cost of the estimation process is high, so computational power is considered a limiting factor in polynomial searches. To improve the performance of these methods, the searched polynomials can be reduced. Thus, there is no longer a need to test all possible polynomials to estimate the generator polynomial of the bit stream.

The statistical test for LFSR estimation operates on the relative percentage of bit stream allocation to Gaussian distribution [53], so the output of these methods identifies the relative percentage of belonging to the Gaussian distribution. The accuracy of the assumed hypothesis depends on the significance value of hypothesis test. The significance value varies depending on the required estimation precision. However, the final answer is also provided as a relative percentage. Therefore, a threshold must be set for a binary decision-making process.

Ultimately, the accuracy of the LFSR pattern discovery method in the input bit stream may not be 100%. The lower the matching percentage, the weaker the reliability of the response, and the higher the matching percentage, the more reliable the percentage is for the existence of an LFSR pattern in the bit stream (Figure 14 , Figure 15).

In non-deterministic artificial intelligence methods, feature extraction is performed on the generated bit stream, and learning models are trained on the produced pseudo-random generator bit stream. In artificial neural networks, feature extraction occurs automatically [56]. In such problems, feature extraction is left to the learning model. Consequently, neural network-based methods have higher computational costs and accuracy compared to other artificial intelligence methods [57]. In these methods, pattern extraction from the output bit stream is performed using neural networks. Depending on the learning method and model, an appropriate function can be fitted to these algebraic relationships [58]. The estimation accuracy on test data can be up to 100% depending on the model's capabilities (Figure 16). However, the accuracy obtained on training data is 100%.

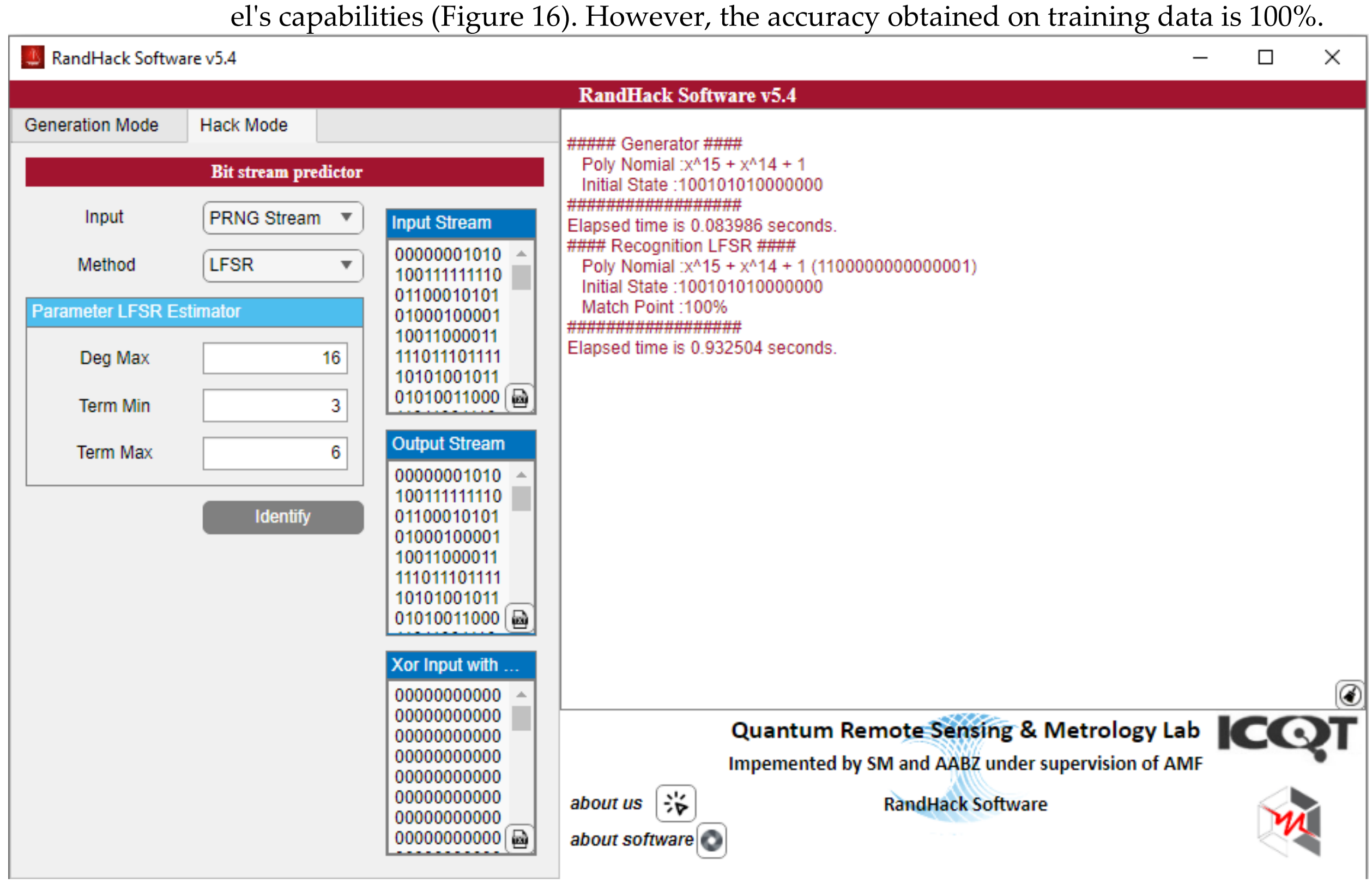

*Figure 14 Breaking the structure of the LFSR pseudo-random sequence using statistical tests in the designed software with specific seed and polynomial*

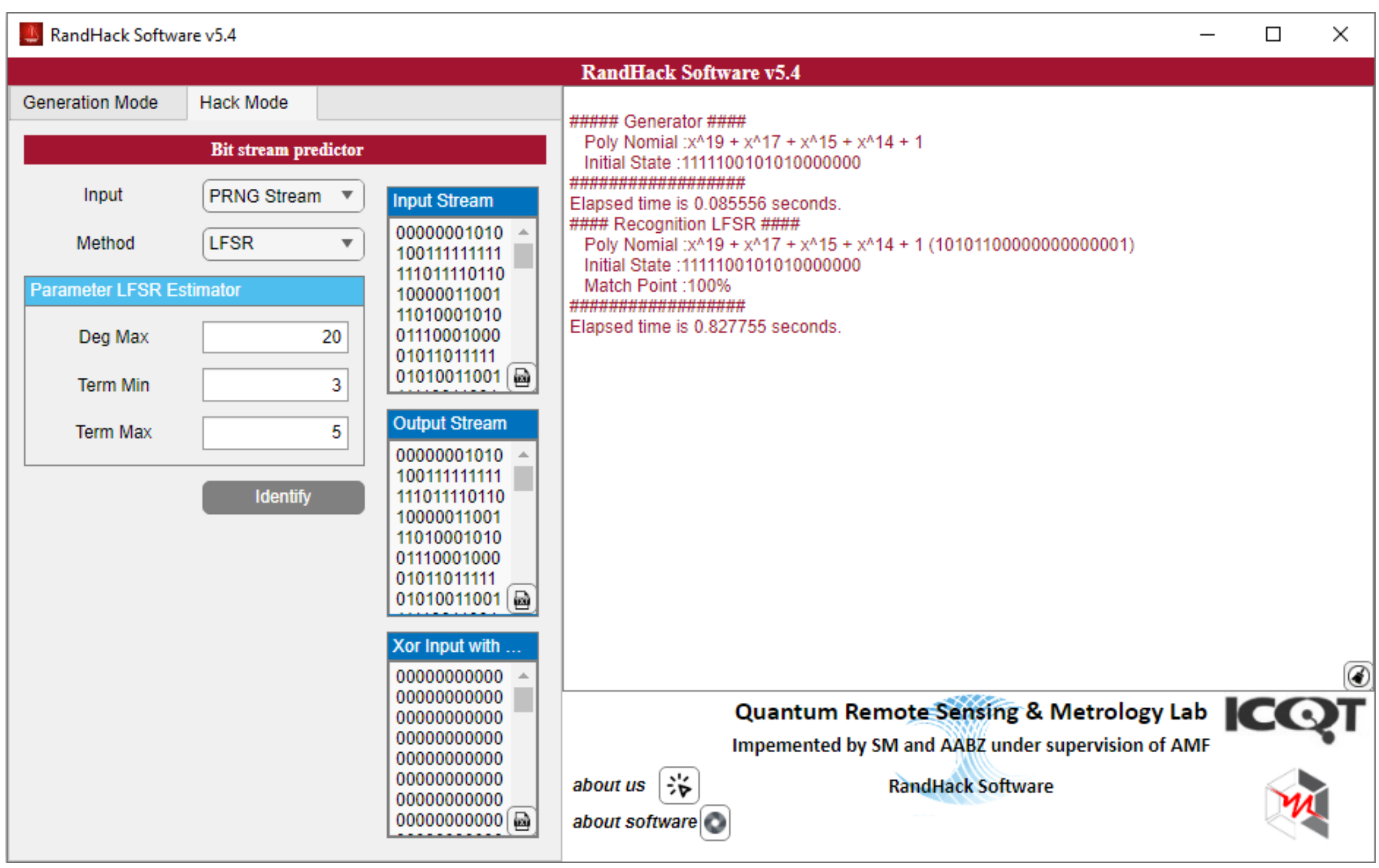

*Figure 15 Breaking the structure of the LFSR pseudo-random sequence using statistical tests in the designed software with specific seed and polynomial*

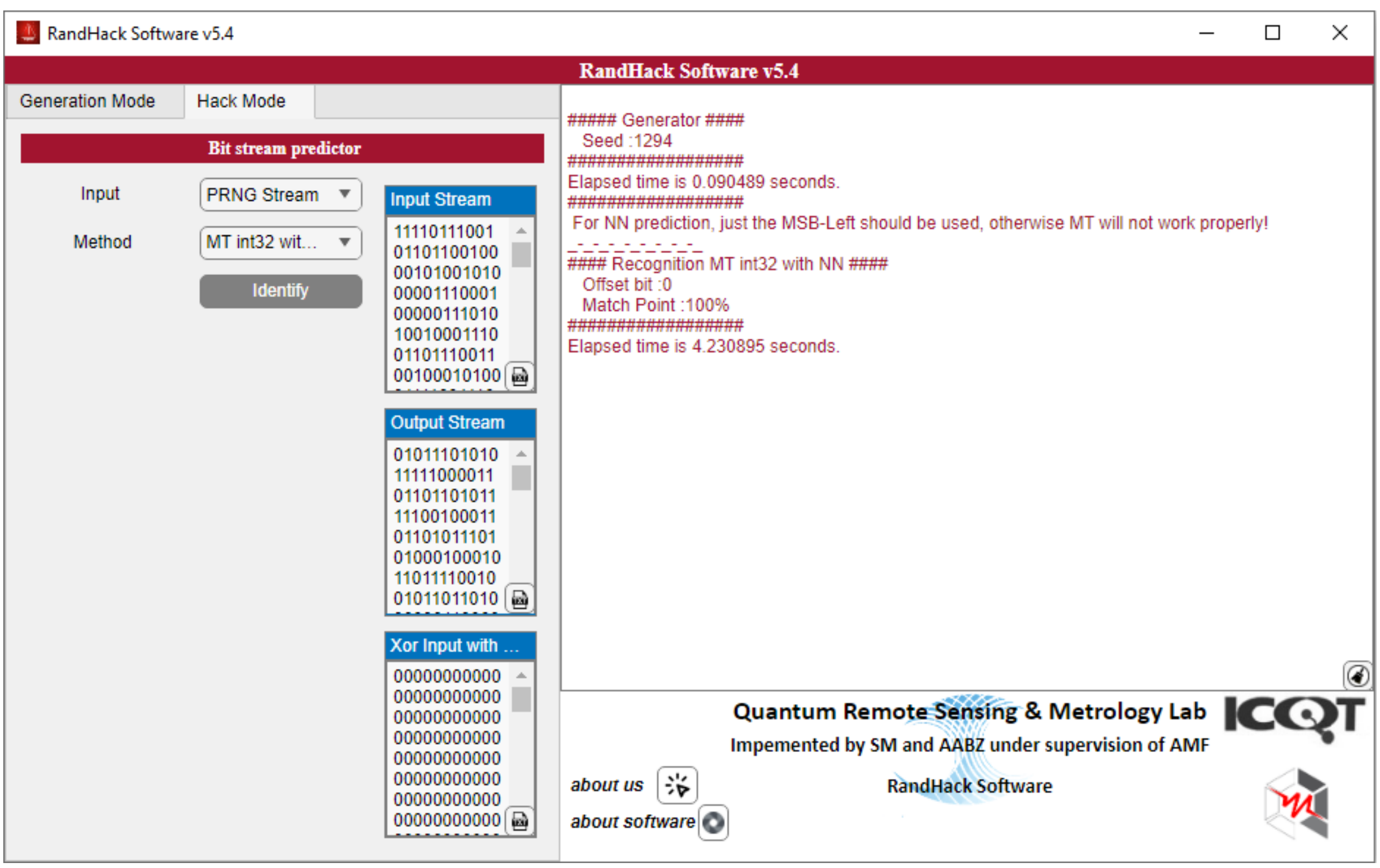

*Figure 16 Breaking the structure of the MT pseudo-random sequence using a neural network in the designed software*

The deep learning model demonstrated promising results in predicting MT sequences, achieving high accuracy after training on a large dataset.

### 4.2.3 MT and LSFR sequences combination prediction

In combined MT and LFSR sequences (also LFSR-LFSR), estimation methods combining deterministic and non-deterministic approaches are used. The estimation method in this case is a combination of previous state methods. However, the order of algorithm extraction is crucial. Since polynomial and initial state estimation of LFSR is less complex depending on the generating polynomial register length, LFSR estimation is performed first, followed by MT estimation. Determining the boundary between MT and LFSR sections in the generated bit stream is extremely important. In the absence of 100% accuracy in the initial LFSR boundary detection, subsequent processes of generator structure estimation and MT bit reproduction will be subject to significant computational challenges and potential error propagation. Therefore, the detection method in this section is a hybrid approach consisting of statistical, deterministic, and algebraic components, with its performance being a composite of LFSR and MT estimation methods (Figure 17 , Figure 18).

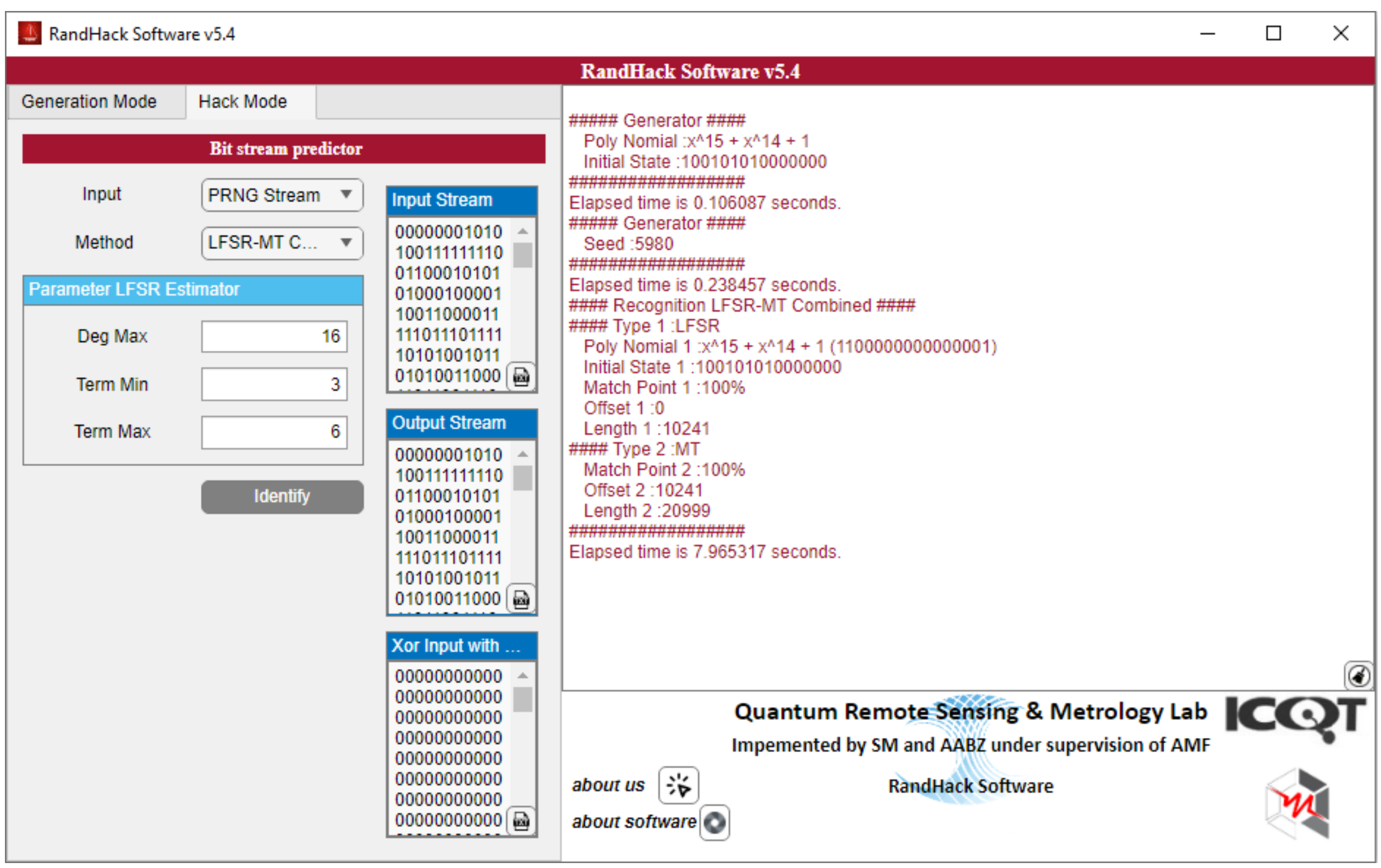

*Figure 17 Breaking the structure of the LFSR-MT pseudo-random sequence using the designed software*

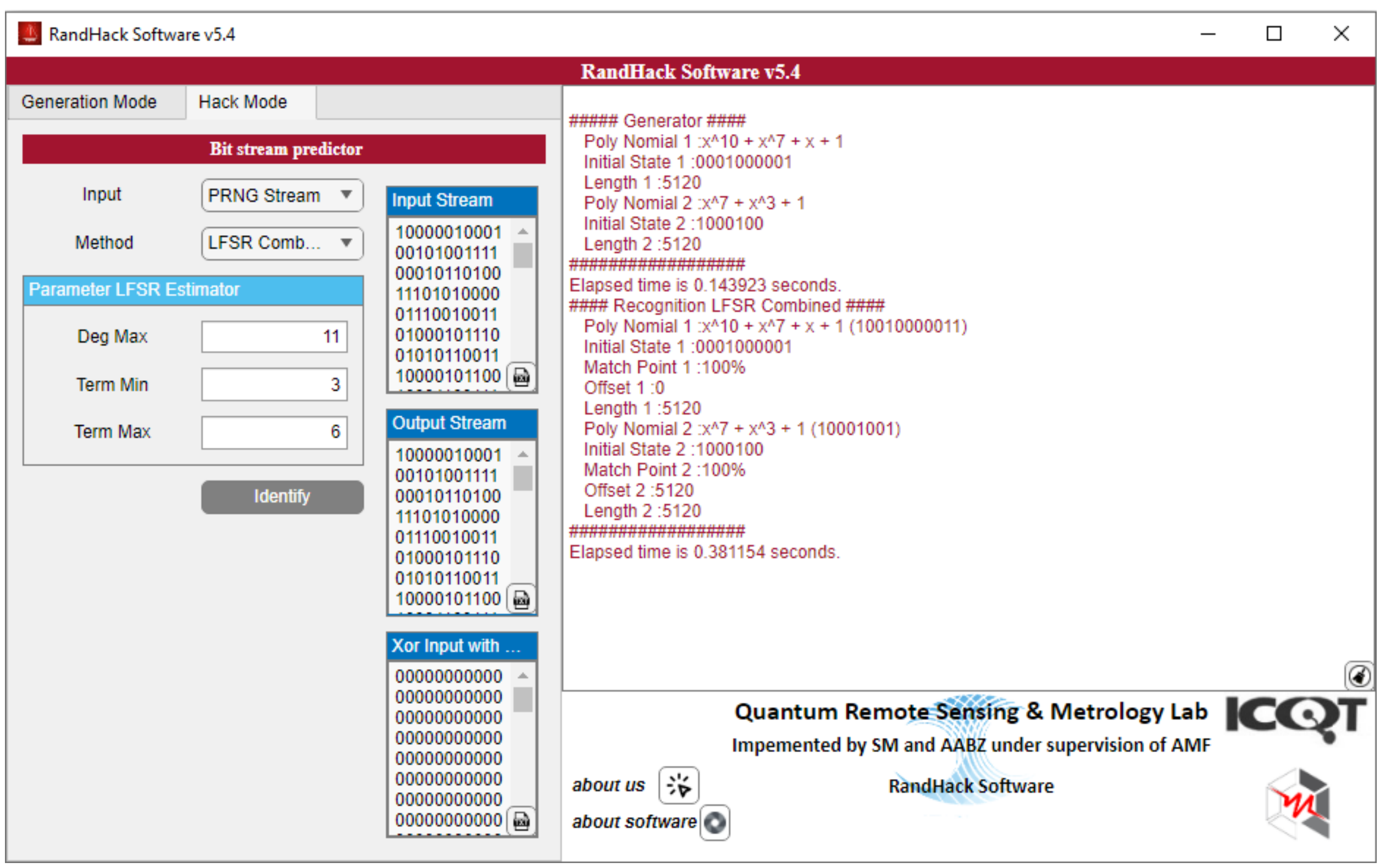

*Figure 18 Breaking the structure of the LFSR-LFSR pseudo-random sequence using the designed software*

4.2.4 Quantum random number sequences case

As an example, to show that quantum random numbers cannot be predicted algorithmically—given their intrinsic randomness [16]—our software provides an option for users to apply LFSR and MT structure estimation algorithms to them. Since quantum random numbers are considered truly random, no pattern or specific trend should exist within the randomly generated data [59], and thus, it can be never hacked even by quantum computers. The implemented estimation algorithms, when run on the quantum random bit stream, are predictably unsuccessful in finding their expected patterns (Figure 19 , Figure 20), as anticipated. Therefore, quantum random numbers demonstrate superiority over a large set of pseudo-random number generation methods based on LFSRs and MT [60] which are among the most practical algorithms in problems requiring random numbers.

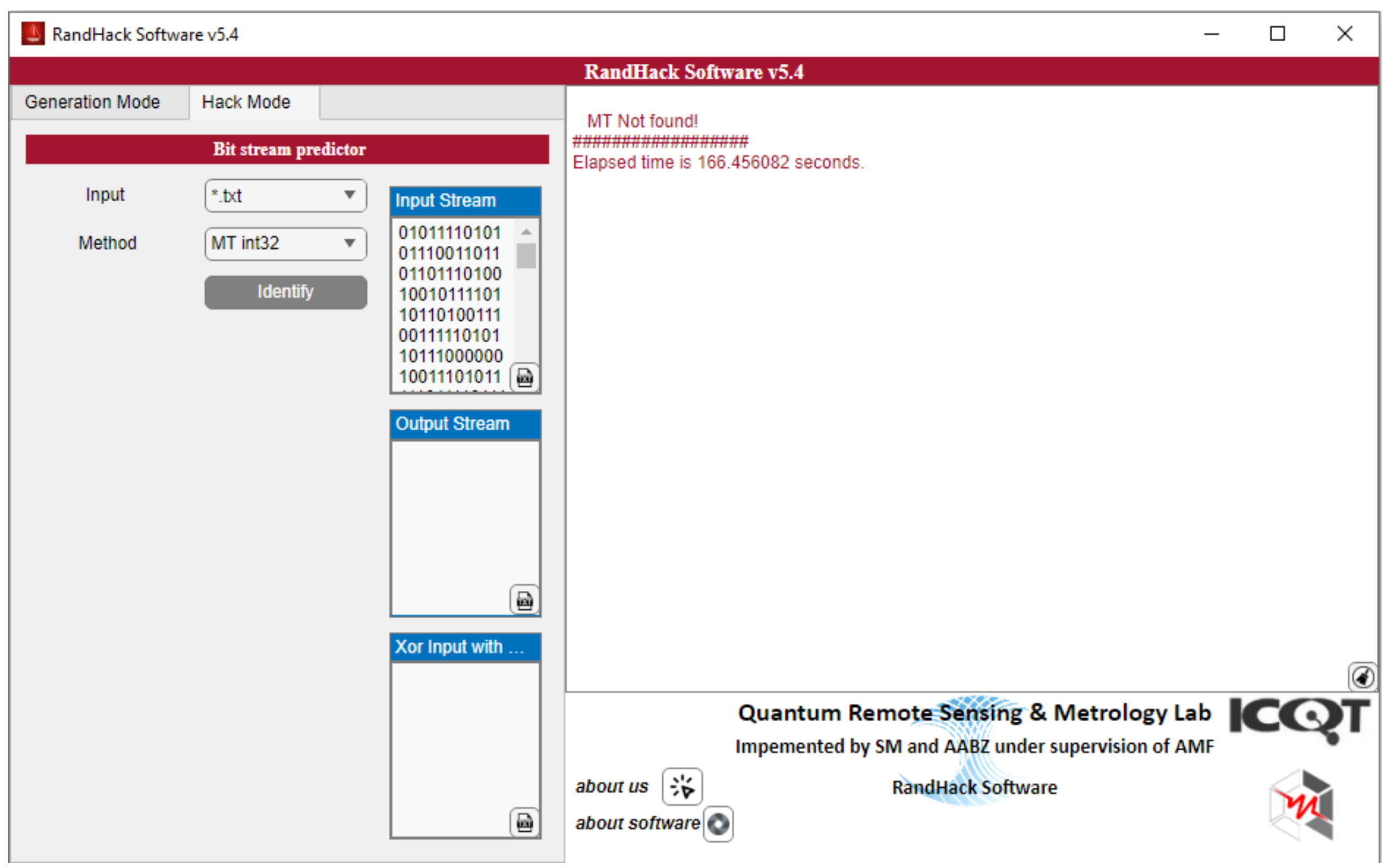

*Figure 19 Identification of the MT pattern in quantum-generated bits using the designed software*

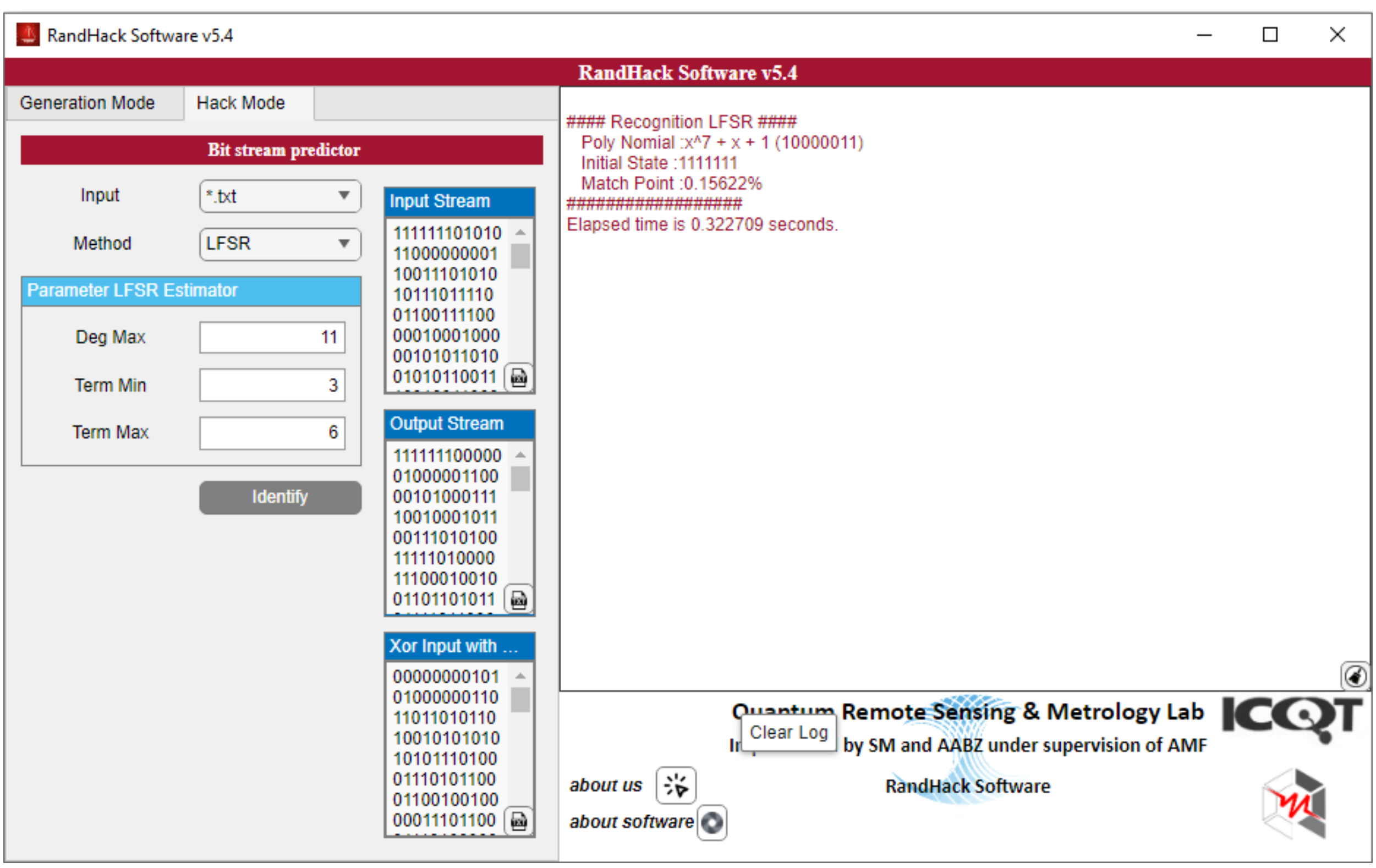

*Figure 20 Identification of the LFSR pattern in quantum-generated bits using the designed software*

## *4.3 Limitation*

### 4.3.1 Computational complexity

In the LFSR structure hacking method, hypothesis testing and search algorithms are used for estimation. Since the used methods are so-called blind methods (meaning no information exists about the polynomial degree or number of structural terms), the complexity of these algorithms increases with higher degrees and larger numbers of terms. For a hardware system with **Error! Reference source not found.** configuration, terms with a maximum degree of 24 provide results at an appropriate scale.

*Table 2 Configuration of the system executing the implemented algorithms*

| **Configuration** | |
|---|---|
| **Processor** | 13th Gen Intel(R) Core (TM) i5-13400 2.50 GHz |
| **Memory (RAM)** | 16.0 GB |
| **Operating System** | Windows 11 Enterprise 24H2 |
| **Simulation Software** | MATLAB 2021b |

Execution time varies based on the length of the bit string checked. A new-generation CPU (like a Core i9) or a powerful GPU (like a 4090 ti) increases processing speed by allowing for the parallel searching of different states. In the MT method, the computational complexity of estimation remains constant, but the volume of calculations for pattern searching along the stream may increase depending on the presence or absence of noise or discontinuity, which is equivalent to searching the entire length of the stream.

## 5. 5. Discussion and Outlooks

Authors should discuss the results and how they can be interpreted from the perspective of previous studies and of the working hypotheses. The findings and their implications should be discussed in the broadest context possible. Future research directions may also be highlighted.

In this section, the limitations and applications of the presented hacking methods are examined. Generally, LFSR generators are used in many applications due to their ease of implementation, speed, and quality of random number generation. These include applications related to security matters where low-correlation bit streams that are quasi-noise are used to generate random sequences. Among the many applications of LFSRs, one can mention the encoded-radar and spread spectrum codes (SSC), where the signal is reduced below the noise level and its power is distributed across different frequencies.

### *5.1 Limitation*

Computational power is one of the major limitations of deterministic methods. Since these methods aim to search for patterns and relationships in the output bit sequence, they have high computational costs when dealing with long output bit sequences. Additionally, in non-deterministic methods, because specific correlations and statistics are extracted from the bit sequence, the computational complexity is again high. Therefore, for applications with high computational complexity and characteristic polynomials, the search range should be determined in accordance with processing power.

Another limitation of these methods is that they operate in the binary field. Mathematical calculations in the binary field have significantly weaker meaning compared to larger fields. As a result, many mathematical operations cannot be performed in the binary field. Implementing the massive MT algorithm on electronic circuits is also costly. Therefore, these methods are suitable for powerful computers, unlike LFSR, which does not have easy implementation. Also, the calculations in these methods are growing exponentially. However, with the development of quantum computers, predictions can be made very quickly.

### *5.2 Application*

Pseudo-random bit sequence generator methods are used in many applications, including the production of random distributions. Random bit sequence distributions are generated, and sampling is conducted on these distributions. Consequently, in random applications, we need to simulate a random state, which is done using pseudo-random bit sequence generators like LFSR pseudo-random generator and MT. The LFSR pseudo-random generator has a significantly higher speed, but the MT method, since it uses twisted LFSR structures to generate random bit sequences, does not have a higher speed compared to LFSR. However, different versions of this method have greater security against hacking attacks compared to LFSR [61]. Accordingly, a primary application of the proposed techniques is the prediction of pseudo-random sequences and the estimate of their internal structure. Hence, whenever an operational random-number generator relies on a LFSR or on the MT, the methods investigated here can be applied effectively. By contrast, these techniques are not suitable for generators that produce truly classical random sequences.

A significant distinction between bit sequences generated by quantum processes and pseudo-random bit sequences lies in their fundamental nature: quantum-generated bit sequences are inherently random, whereas pseudo-random bit sequences follow algorithmic patterns for generation, rendering them vulnerable to breaking structure with increased computational power or by quantum computers. For instance, the implementation of Grover's algorithm on quantum computers reduces the computational complexity required to compromise AES encryption by half, resulting in a complexity order of $2^{128}$for 256-bit key length [62]. This phenomenon illustrates that the computational complexity associated with predicting and deterministically analyzing AES depends on technological advancements and the discovery of novel complexity-reducing algorithms. Conversely, quantum-generated bit sequences remain impervious to such vulnerabilities since the inherently random nature of quantum phenomena precludes prediction even with quantum computers possessing substantial parallel computation capabilities [63]. Although there's a possibility of side-channel attacks like those in cryptography for hacking the quantum random number generators introduced in [64].

### *5.3 Jamming*

Jammers are devices employed to disrupt communication between multiple nodes. Various jammer configurations exist, capable of disrupting either narrow frequency spectrum segments or entire frequency bands. When narrow-band jammers are deployed, power consumption remains relatively low; however, these jammers can be circumvented through carrier frequency modulation at the transmitter. Conversely, when wideband jammers are utilized to disrupt an extensive frequency range, they require significant power consumption for signal transmission [65].

An efficient countermeasure to frequency-hopping communication is to infer the hopping pattern and deploy a targeted jamming signal. Conventional frequency-hopping schemes typically derive their hop sequences from pseudo-noise patterns produced by LFSRs. Once the underlying order of these patterns is estimated, the hopping process can be neutralized [53].

As evidenced by the techniques implemented in our software, hop patterns based on pseudo-random sequences become predictable upon the reconstruction of the generator's structure and initial state, thereby making their jamming straightforward. Thus, the idea of a quantum hopping or quantum jammer can be put forward. when truly random sequences are used to drive the hopping pattern, predicting future hops becomes extremely challenging—and in some cases infeasible. Consequently, employing quantum random sequences, whose stochasticity originates in fundamental quantum phenomena, greatly complicates any attempt to jam a frequency-hopping system that follows a quantum-based hopping pattern [66].

## 6. Summary and Conclusions

In conclusion, we developed a user-friendly software, which systematically explored the predictability of pseudo-random sequences generated by LFSR, MT, and their hybrid combinations, conclusively demonstrating inherent vulnerabilities in these widely used classical algorithms due to their deterministic nature. The successful reconstruction of LFSR parameters and the accurate prediction of MT sequences, even with deep learning models achieving perfect accuracy by exploiting latent temporal correlations and deterministic operations, underscore these limitations. While, quantum-generated sequences have proven impervious to such prediction techniques, firmly establishing the superiority of true randomness for high-security applications. The software enables users to generate LFSR/MT and predict all of them, as well as to get a result file.

Key insights reveal the fundamental weaknesses of deterministic/classical generators, their susceptibility to exploitation by machine learning models, and the inherent resilience of quantum-derived randomness. While hybrid approaches aimed to enhance complexity offer some resistance, their underlying linear foundations remain exploitable. The practical implications of these findings are significant for various security-critical domains, highlighting the risks associated with relying on predictable pseudo-randomness.

Future research should investigate the resistance of the enhanced-PRNGs, that integrate physical randomness with algorithmic efficiency, against quantum computers. On the other hands, beyond the use of QRNG in order to improve PRNGs it is needed to explore advanced and adaptive deep learning architectures for preemptively identifying vulnerabilities, and design non-linear hybrid generators to balance security and performance. As computational and adversarial capabilities continue to advance, the transition towards quantum-secure methods for random number generation will become increasingly imperative, ensuring robust randomness as a cornerstone of next-generation cryptographic, communication, and other critical systems. Moreover, we briefly introduce the idea of quantum jamming/hopping in which the quantum random number/noise can be used instead of PRNG in order to be robust against hack. Finally, the present work provides a crucial picture for advancing and importance of secure random number generation and its predicting methods in an era defined by escalating computational threats.

**Author Contributions:** AMF defined and lead the project as supervisor and group leader of quantum sensing and metrology at ICQT. AAZ have performed and implemented the algorithms as well as the GUI software. The manuscript has been written by AAZ and edited by AMF. Both authors contribute to prepare the manuscript.

**Acknowledgments:**

The authors would like to thank Iranian Centers for Quantum Technology (ICQT) for their support. AAZ also thank Eng. MS who helps him.

**Conflicts of Interest:** The authors declare no conflicts of interest.

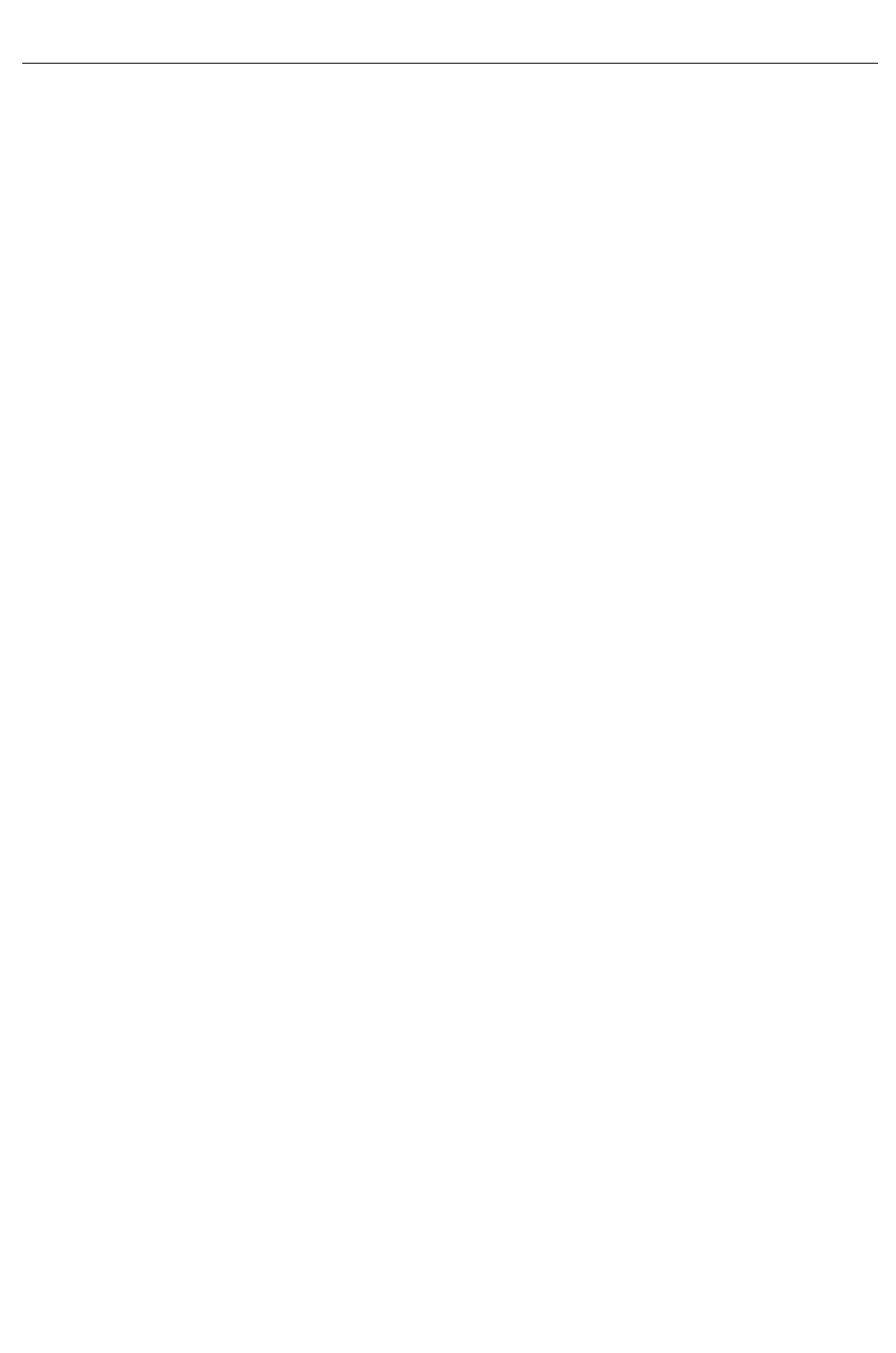